\documentclass[aps,prr,reprint]{revtex4-1}
\usepackage{epsfig}
\usepackage{dcolumn}
\usepackage{physics}
\usepackage{graphicx,graphics}
\usepackage{amsmath}
\usepackage{amssymb}
\usepackage{bm}
\usepackage{color}
\usepackage[utf8]{inputenc}
\usepackage[colorlinks,linkcolor={blue},citecolor={blue},urlcolor={blue}]{hyperref}
\usepackage{mathtools}
\usepackage{booktabs}
\usepackage{subfigure}

\newcommand{\Giless}{ {\rm G}^{ <}}

\newcommand{\bk}{{\bm k}}

\newcommand{\BJ}{\mathcal{J}_0}
\newcommand{\BY}{\mathcal{Y}_0}
\newcommand{\BK}{\mathcal{K}_0}
\newcommand{\BI}{\mathcal{I}_0}

\DeclareMathAlphabet{\mathpzc}{OT1}{pzc}{m}{it}

\begin{document}

\title{Non-Equilibrium RKKY Interaction in Irradiated Graphene}

\author{Modi Ke}
\thanks{These two authors contributed equally.}
\affiliation{Department of Physics and Astronomy, The University of Alabama, Tuscaloosa, AL 35487, USA }
\author{Mahmoud M. Asmar}
\thanks{These two authors contributed equally.}
\affiliation{Department of Physics and Astronomy, The University of Alabama, Tuscaloosa, AL 35487, USA }
\author{Wang-Kong Tse }
\affiliation{Department of Physics and Astronomy, The University of Alabama, Tuscaloosa, AL 35487, USA }

\begin{abstract}
We demonstrate that the Ruderman-Kittel-Kasuya-Yosida (RKKY) interaction in graphene can be strongly modified by a time-periodic driving field even in the weak drive regime. This effect is due to the opening of a dynamical band gap at the Dirac points when graphene is exposed to circularly polarized light. Using Keldysh-Floquet Green's functions, we develop a theoretical framework to calculate the time-averaged RKKY coupling under weak periodic drives and show that its magnitude in undoped graphene can be decreased controllably by increasing the driving strength, while mostly maintaining its ferromagnetic or antiferromagnetic character. In doped graphene, we find RKKY oscillations with a period that is tunable by the driving field. When a sufficiently strong drive is turned on that brings the Fermi level completely within the dynamically opened gap, the behavior of the RKKY coupling changes qualitatively from that of doped to undoped irradiated graphene.
\end{abstract}

\maketitle

\section{Introduction} \label{intro}

The Ruderman-Kittel-Kasuya-Yosida (RKKY) coupling is the indirect exchange interaction in a magnetically doped system, in which the coupling between the localized impurity spins is mediated via the conduction electrons of the host material~\cite{RKKY1,RKKY2,RKKY3}.
As a function of the impurity separation, it oscillates between positive and negative values indicating ferromagnetic and antiferromagnetic spin couplings with a period determined by the host's Fermi level~\cite{kittelbook}. The envelop of these oscillations decays in a power law that depends on the dimensionality and the specific band structure of the host material~\cite{serami,blackshaffer,satpathy1,satpathy2,GFR1,GFR2,MOS2RKKY3,RKKYRashbaSem}. For systems with parabolic bands, the envelope of the RKKY oscillations decays as $R^{-3}$ in three dimension (3D) and as $R^{-2}$ in two dimension (2D)~\cite{RKKY1,RKKY2,RKKY3}.

The 2D nature and Dirac energy dispersion of graphene have spawned the discovery of many unique and intriguing electronic properties~\cite{graphene1,graphene2}. In particular, the RKKY interaction between magnetic impurities deposited on graphene exhibits peculiar properties that are quite different from conventional systems. For undoped graphene, the power law of the RKKY oscillations is $R^{-3}$ in contrast to $R^{-2}$ expected in 2D. In addition, the RKKY oscillations do not alternate between ferromagnetic and antiferromagnetic values but can instead display short-ranged fluctuations depending on the locations of the impurity atoms on the honeycomb lattice~\cite{serami,blackshaffer,satpathy1,satpathy2,Power_review}. This dramatic departure from the normal 2D RKKY behavior is not unexpected since the RKKY oscillation is fundamentally a Fermi surface phenomenon, and undoped graphene is characterized by a pair of Fermi points at the Dirac nodes. In doped graphene, some of the conventional 2D RKKY behavior is recovered due to the presence of a finite Fermi surface. Hence, on top of possible short-ranged fluctuations due to the lattice, its RKKY interaction oscillates between ferromagnetic and antiferromagnetic values, with the envelope decreasing as $R^{-2}$. This makes the RKKY coupling of doped graphene qualitatively similar to that of a 2D systems with a parabolic dispersion~\cite{satpathy2,GFR2}.

The long-range nature of the indirect exchange interaction provides an important mechanism in layered magnetic structures allowing for the controlled storage and transfer of spin-encoded information~\cite{layered1,layered3}. The capability to further manipulate the indirect exchange interaction would add another dimension towards the possibility of engineering new material behaviors and functionalities. Over the past two decades, there has been much interest and progress in the studies of ultrafast optical control of spin dynamics and magnetic order \cite{ultrafast1}.  An  emerging strategy to achieve this end in magnetically ordered materials is through optical control of direct exchange coupling \cite{optex1,optex2}. In magnetically doped or layered systems, RKKY interaction is an important type of exchange coupling to consider. Optical manipulation of RKKY interaction was first studied in semiconductor quantum dot systems and II-VI diluted magnetic semiconductors~\cite{optRKKY1,optRKKY2,optRKKY3,optRKKY4}, where the exchange interaction can be mediated by virtual electron-hole pairs or excitons generated by a sub-band gap optical excitation. Manipulation of RKKY interaction is also possible through materials engineering of magnetic multilayers, where the interlayer interaction is predominantly due to RKKY coupling~\cite{Bruno_review,Stiles2005}. An applied static electric field~\cite{Arun_paper,HChen,RKKYvolt1,RKKYvolt4,blgcont} could also provide a means to control the RKKY interaction in materials and structures with considerable spin-orbit coupling~\cite{MOS2RKKY3,MOS2RKKY2,RKKYRashbaSem,RKKYSOC}.

Recently, there has been a burgeoning interest in the study of periodically driven Floquet systems and the tantalizing possibility to engineer and control such non-equilibrium quantum  states~\cite{OkaFloqEngRev}.
Periodic driving through monochromatic irradiation of a solid renormalizes the band structure through non-perturbative light-matter coupling into Floquet-Bloch bands. By tuning the intensity and frequency of the radiation one can dynamically tune these photon-dressed bands and hence control the properties of the irradiated system. Besides the control of direct exchange coupling already mentioned, Floquet dynamics also strongly influence transport properties~\cite{fg1,Floqtrans5,Floqtrans4}, optical properties~\cite{Floqopt1,Floqopt2,Floqtrans3}, topological  phases~\cite{Floqtop2,Floqtop4,RudnerRev} and interlayer tunneling~\cite{Platero,FloqTunn4}.

The dynamic tunability and exquisite level of control offered by optical driving can provide a significant advantage over other proposed mechanisms for controlling the RKKY interaction. In this work, we study  the non-equilibrium RKKY interaction in magnetically doped graphene under illumination of monochromatic circularly polarized light. Due to their linear dispersion and valley degrees of freedom, chiral Dirac fermions in graphene couple distinctively to electromagnetic field compared to Schr\"{o}dinger fermions with a parabolic dispersion. Indirect exchange interaction mediated by \textit{irradiated} Dirac fermions are therefore expected to display unconventional properties not found in equilibrium.
Our theory is developed using the Keldysh-Floquet formalism in order to capture the non-perturbative Floquet dynamics of the RKKY interaction.
We derive a formula for the time-averaged RKKY coupling for the weak drive, off-resonant regime, and perform analytical and numerical studies as a function of  impurity locations on the lattice sites and Fermi levels, for various driving strengths.
Our results reveal that for undoped graphene a weak periodic driving primarily decreases the RKKY exchange interaction in proportion to the driving strength, whereas for doped graphene increasing the driving strength increases the period of RKKY oscillation allowing for a optical tuning of indirect exchange coupling. We are also able to identify a parameter regime where the exchange coupling can be tuned between ferromagnetic and antiferromagnetic.

Our paper is organized as follows. In Sec.~\ref{System and the method} we introduce the theoretical model of graphene with magnetic adatoms under monochromatic illumination.  We derive the Floquet Hamiltonian for irradiated graphene and then use the Keldysh-Floquet formalism to derive a general formula for the time-averaged non-equilibrium RKKY interaction. We then specialize to the weak drive limit and obtain an approximate Floquet Hamiltonian under the photon-number representation and the corresponding expression of the irradiated RKKY coupling.  We next proceed to obtain  the explicit analytical formulas of real-space non-equilibrium Green's functions in Sec.~\ref{Calculation of the Green's Funtions}. In Sec.~\ref{Numerical results} we present and analyze  our results for the time-averaged RKKY coupling for the cases of undoped graphene (Sec.~\ref{intrinsic}) and doped graphene (Sec.~\ref{doped}) under different impurity configurations. These numerical results are elucidated with analytical studies where we obtain approximate analytic expressions for the irradiated RKKY coupling and exhibit their dominant dependence on the impurity separation. Before concluding in Sec.~\ref{conclusion}, we present a few relevant remarks on our results and an outlook for future direction in Sec.~\ref{Discussion}.

\section{Formulation} \label{System and the method}

\begin{figure}
    \begin{center}
            \includegraphics[width=1\columnwidth]{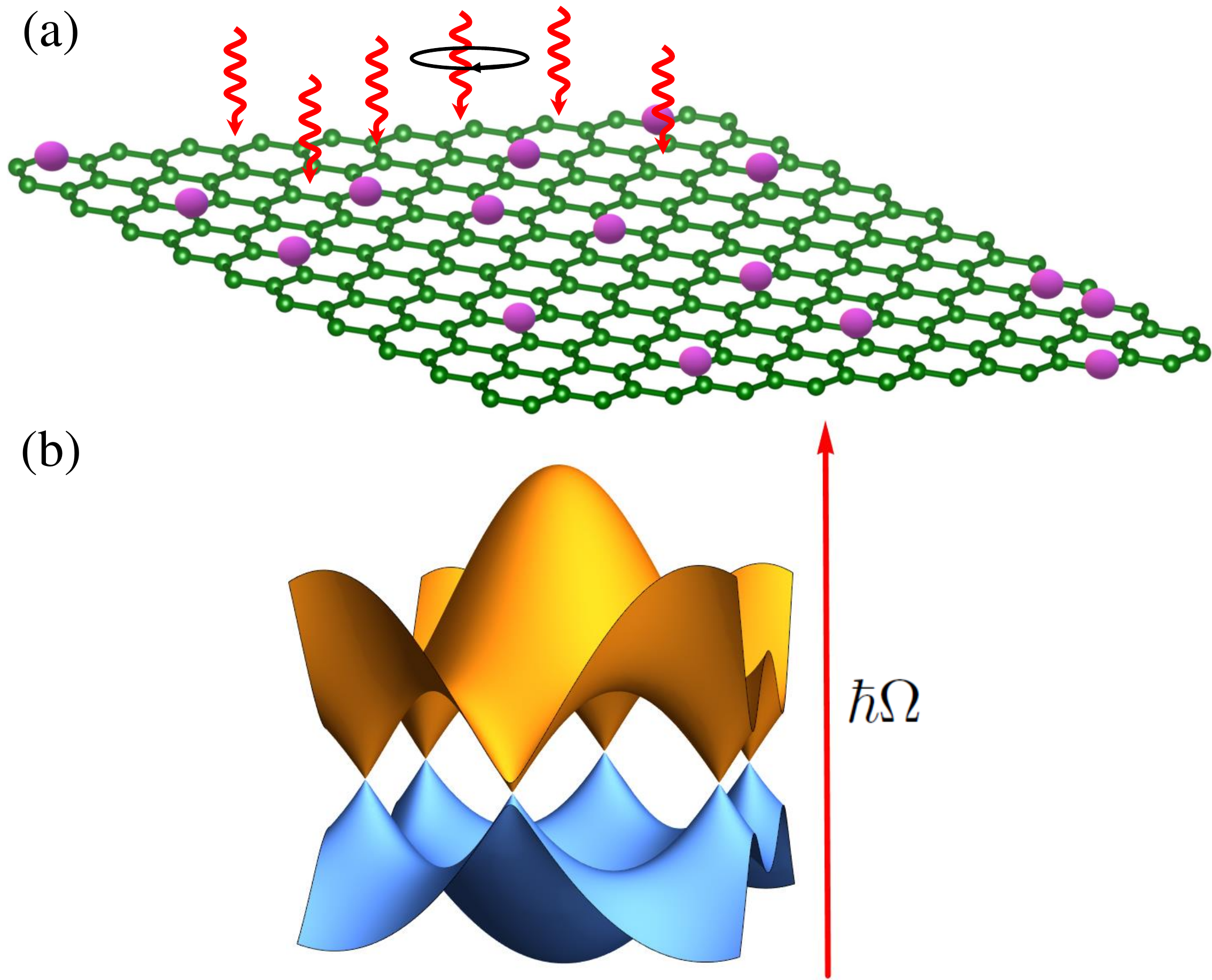}
                \end{center}
                \caption{(a) Schematic of irradiated graphene sheet deposited with magnetic impurity adatoms. (b) Tight-binding dispersions of the $\pi$-bands of graphene~\cite{graphene1}. The system is driven in the off-resonant regime at a frequency larger than the band width $\hbar\Omega> 6\gamma$.  }
    \label{newfig1}
\end{figure}

The low-energy electrons in single layer graphene are described by two inequivalent high-symmetry points ($K$ and $K'$) in the Brillouin zone known as the Dirac points. Electrons in the vicinity of  these points disperse linearly and can be described by a Dirac Hamiltonian~\cite{graphene1,graphene2}
\begin{eqnarray}\label{h0tau}
\mathcal{H}_0=\hbar v_F(\tau \sigma_x k_x+ \sigma_y k_y),
\end{eqnarray}
where $v_{F}=10^{6}$ ms$^{-1}$ is the band velocity and $\tau =1$ $(\tau=-1)$ corresponds to the $K$ $(K')$ Dirac points. $v_{F}$ is related to the graphene tight-binding model parameters by $\hbar v_{F}=3\gamma a/2=6.6\;\mathrm{eV\AA}$ with $\gamma=3$ eV being the nearest-neighbor hopping amplitude and $a=1.4$ \AA\; the carbon-carbon distance.
In the presence of magnetic impurity adatoms, electrons in graphene couple with the impurity spins through a short-range interaction. In the spirit of standard RKKY theory, such an interaction is  modeled as a local potential at the impurity site and the single-particle Hamiltonian takes the form
\begin{eqnarray}
\mathcal{H} = \mathcal{H}_{0}+\lambda \delta ( \boldsymbol{r} )  \boldsymbol{S}  \cdot\boldsymbol{\mathbb{S}}_1+\lambda \delta ( \boldsymbol{r} - \boldsymbol{R} )  \boldsymbol{S}  \cdot\boldsymbol{\mathbb{S}}_2,
\end{eqnarray}
where ${\bm S}$ is the electron spin operator, $\boldsymbol{\mathbb{S}}_i$ ($i = 1,2$) are the spin angular momenta of the magnetic impurities assumed to be located at the origin and at ${\bm R}$, and $\lambda$ is the coupling between the itinerant and localized spins. The honeycomb graphene lattice can be considered as two superposing sublattices $A$ and $B$. The location of the two impurity spins can be either at the sites of the same sublattice $A$ or different sublattices $A$ and $B$.

We consider normal incidence on the graphene plane with a circularly polarized (CP) light ${\bm E}=E_0[\cos{(\Omega t)}\hat x+\sin{(\Omega t)}\hat y]/\sqrt{2}$, with frequency $\Omega$ and field amplitude  $E_{0}$ [see Fig.~\ref{newfig1}(a)]. The electric field couples to the Hamiltonian in Eq.~\eqref{h0tau} via the minimal coupling scheme and the resulting time-dependent Hamiltonian of the irradiated graphene system becomes
\begin{eqnarray}\label{tdph}
  \mathcal{H}_0(t) &=& \hbar v_F \tau \sigma_x k_x+\hbar v_F  \sigma_y k_y\nonumber
\\
&&-A \tau \sigma_x \sin{(\Omega t)}+A\sigma_y \cos{(\Omega t)},
\end{eqnarray}
where $A= v_F e E_0/(\sqrt{2}\Omega)$.

Due to the time-periodicity of the Hamiltonian Eq.~\eqref{tdph}, the Floquet-Bloch theorem is satisfied granting a solution of the following form to the time-dependent Dirac equation $\mathcal{H}_0(t)\left|\psi(t)\right\rangle = i\hbar\partial_t \left|\psi(t)\right\rangle$:
\begin{equation}
\left|\psi_{l,\bk}(t)\right\rangle=e^{-i \epsilon_{l,\bk}t/\hbar}\left|u_{l,\bk}(t)\right\rangle,
\end{equation}
where $l\in Z$ labels the Floquet modes and $\epsilon_{l,\bk}$ is the quasienergy modulo $\hbar\Omega$. The time periodic nature of the Floquet states, $\left|u_{l,\bk}(t+T)\right\rangle=\left|u_{l,\bk}(t)\right\rangle$, where $T=2\pi/\Omega$ is the driving period, further allows a Fourier series representation of these states,
\begin{equation}
\left|u_{l,\bk}(t)\right\rangle=\sum_{n}e^{-i n \Omega t}\left|u_{l,\bk}^{n}\right\rangle.
\end{equation}
Substituting $\left|\psi_{l,\bk}(t)\right\rangle$ in Eq.~\eqref{tdph} maps the time-dependent Dirac equation into the Floquet Hamiltonian, $\mathcal{H}_{F}=\mathcal{H}_0(t)-i\hbar\partial_{t}$ in the Fourier domain~\cite{shirley}, such that
\begin{equation}\label{floqh}
\sum_n(\mathcal{H}_{mn}-n\hbar \Omega \delta_{m,n})|u_{l,\bk}^{n}\rangle=\epsilon_{l,\bk}|u_{l,\bk}^{m}\rangle,
\end{equation}
where $\mathcal{H}_{F,mn}=\mathcal{H}_{mn}-n\hbar \Omega \delta_{m,n}$ is the Floquet Hamiltonian and
\begin{equation}
\mathcal{H}_{mn}=\frac{1}{T}\int^{T}_{0}dt e^{i(m-n)\Omega t}\mathcal{H}_{0}(t),
\end{equation}
is known as the Floquet matrix~\cite{Folqthe,Oka_RMP}. For our system, the Floquet matrix takes a block-tridiagonal form and can be explicitly written as
\begin{eqnarray}\label{flm}
\mathcal{H}_{mn}&=&\hbar v_F(\tau \sigma_x k_x+ \sigma_y k_y)\delta_{mn}+\frac{i}{2}A[(\tau \sigma_x-i\sigma_y)\delta_{m,n-1}\nonumber
\\
&-&(\tau \sigma_x+i\sigma_y)\delta_{m,n+1}].
\end{eqnarray}
where $\tau=+$ ($\tau=-$) corresponds to the $K$ ($K'$) point.  Here we should note that the adoption of the continuum description of graphene is essential to obtaining the block-tridiagonal form of the Floquet Hamiltonian of irradiated graphene. The adoption of this description, however, sets an upper bound to the light-matter coupling strength $\mathcal{A}=A/(\hbar \Omega)$, where $A$ is given in Eq.~\eqref{tdph}. This upper bound can be seen from the following consideration, starting first with the tight-binding model of graphene. The Floquet Hamiltonian can be derived from this tight-binding description, which consists of a Floquet matrix with every elements being generally non-zero. If we require this to recover the continuum description of the Floquet Hamiltonian in Eq.~\eqref{flm}, all elements of the block-pentadiagonal and higer-ordered bands of this matrix must be set to zero (here the terminology ``bands'' refer to the bands of the matrix). The matrix elements in those bands are proportional to $\mathcal{J}_{n\geqslant 2}[\sqrt{2}a A/(\hbar v_{F})]$, where $\mathcal{J}_{n}$ is the Bessel function of order $n$. The condition $\mathcal{J}_{n\geqslant 2}[\sqrt{2}a A/(\hbar v_{F})]\approx0$ gives $a A/\hbar v_{F}\ll 2$ and thus $\mathcal{A} \ll 3\gamma/\hbar\Omega$. Here $a$ and $\gamma$ are defined under Eq.~\eqref{h0tau}.
For our numerical calculations in Sec.~\ref{Numerical results}, we choose $\hbar \Omega = 6.6\gamma$, which puts the driving field in off-resonance [see Fig.~\ref{newfig1}(b)], and $\mathcal{A} \leqslant 0.06$ so that both the above condition and the weak drive condition are satisfied with $\mathcal{A} \ll 3\gamma/\hbar\Omega < 1$.
The above consideration was first discussed in the context of irradiated two-dimensional electron gases in Ref.~\cite{AsmTse} and can be seen to apply quite generally for other systems whenever one tries to obtain a Floquet Hamiltonian within a continuum low-energy description.

Turning on the driving field at $t = 0$ starts pumping energy into the system. Crucially, the electron subsystem in a solid is always an open system, with energy relaxation pathways through various inelastic scattering processes (electron-electron and electron-phonon scattering) as well as coupling to extrinsic degrees of freedom in the ambient environment.
In the presence of both energy injection and relaxation, naturally heat does not build up infinitely over time. We adopt a specific model to account for energy relaxation by coupling graphene to an external heat reservoir, which will be specified in more details in a later section, Sec.~\ref{Calculation of the Green's Funtions}.
After initial transients has subsided, the system dynamics settles into a non-equilibrium steady state (NESS). In NESS, time periodicity is restored and the system's Green's functions become periodic in time, $G^{{\rm R,A,<}}_\bk(\bm r,t+T;\bm r',t'+T)=G^{{\rm R,A,}<}_\bk(\bm r, t;\bm r',t')$, where $G^{{\rm R,A,}<}$ denotes the retarded, advanced, and lesser Green's function.
Periodicity allows these Green's functions to be expressed in the Floquet representation \cite{Oka_RMP},
\begin{eqnarray}
[G(\bm r,\bm r',\bar{\omega})]_{mn}&=&
\\
\frac{1}{T}\int^T_0 d t_{\textrm{av}}\int^{\infty}_{-\infty}&d t_{\textrm{rel}}& e^{i(\bar{\omega}+m\Omega)t-i(\bar{\omega}+n\Omega)t'}G(\bm r,t;\bm r',t'), \nonumber
\end{eqnarray}
where $t_{\textrm{av}}=(t+t')/2$ and $t_{\textrm{rel}}=t'-t$ are the average time and relative time. The frequency variable $\bar{\omega}$ in the above is defined in the reduced zone $\bar{\omega} \in (-\Omega/2,\Omega/2]$. We will use the notation $\bar{\omega}$ for reduced-zone frequency and  $\omega \in (-\infty,\infty)$ for extended-zone frequency, which will be relevant for the Green's functions we will discuss further below.

The RKKY interaction coupling is mediated by the spin polarization of the graphene electrons induced by the impurity spins. We use perturbation theory to obtain the interaction energy up to leading order in $\mathbb{S}_{1}\cdot\mathbb{S}_{2}$. The exchange energy between impurity spins $\mathbb{S}_{1}$ and $\mathbb{S}_{2}$ is given by the interaction energy between one impurity spin and the induced electron spin density due to the other impurity spin,
\begin{eqnarray}\label{exch}
E({\bm R},t)=\lambda \int d{\bm r} \mathbb{S}_{2,\mu} \delta ({\bm r}-{\bm R})  \langle\delta{S}_{\mu}({\bm r},t)\rangle,
\end{eqnarray}
where $\langle{\bm S}({\bm r},t)\rangle=-i{\rm Tr}[{\bm S}\delta
\Giless ({\bm r},t;{\bm r},t) ]$ is the expectation value of the
induced spin density, with $\delta\Giless ({\bm r},t;{\bm r},t)$ denoting the change in $\Giless$ due to the perturbation of  $\mathbb{S}_{1}$. Dyson's equation up to first order in
the local  potential $V(\bm r)=\lambda \delta ( {\bm r} )  {\bm S}  \cdot { \mathbb{S}}_1$ gives
 \begin{eqnarray}\label{Dyson}
\langle\delta{\bm S}(\bm r,t)\rangle &&= - i{\rm Tr} \{ {\bm S} G^{\rm R}(\bm r,t;\bm r',t') V(\bm r') G^<(\bm r',t';\bm r,t)\nonumber\\ &&+ {\bm S}  G^<(\bm r,t;\bm r',t') V(\bm r') G^{\rm A}(\bm r',t';\bm r,t)  \},
\end{eqnarray}
where $G^{{\rm R,A},<}$ are the non-interacting Green's functions of the irradiated graphene sheet, the trace is over $\bm r'$,  $t'$  and the spin
degrees of freedom. Let the impurity spin $\mathbb{S}_{1}$ be sitting on the sublattice site $\beta$ and $\mathbb{S}_{2}$ be sitting on $\alpha$, where $\alpha,\beta = \{A,B\}$.
Then, substituting Eq.~\eqref{Dyson} in Eq.~\eqref{exch}, expressing the time-dependent Green's functions in the Floquet representation and averaging over time,
we obtain the Floquet representation of the time-averaged exchange energy for impurity spins located at sublattice sites $\alpha$ and $\beta$ (for details see Appendix~\ref{apexchf0}):
\begin{eqnarray}\label{exch1}
&E&_{\alpha \beta}({\bm R})=- i \lambda^2 \sum_{\mu,\nu, a,b} (S_{\mu})_{a b}  (S_{\nu})_{b a} \mathbb{S}_{1,\mu} \mathbb{S}_{2,\nu}\nonumber
 \\
  &\times&\int^{\frac{\hbar\Omega}{2}}_{-\frac{\hbar\Omega}{2}} \frac{d\bar{\omega} }{2\pi}  \textrm{Tr}\{G^{\rm R}_{\alpha \beta}({\bm R},\bar{\omega})G^<_{ \beta\alpha}(-\boldsymbol{R},\bar{\omega})\\
&+&G^<_{\alpha \beta}({\bm R},\bar{\omega})G^{\rm A}_{\beta\alpha }(-{\bm R},\bar{\omega})\},\nonumber
 \end{eqnarray}
 where $\mu,\nu$ are the $x,y,z$-projections of the impurity spin, $a,b$ label the electron spin degrees of freedom, $\bar{\omega}$ is the reduced-zone frequency, and the trace here is over the Floquet index of the Floquet Green's functions. In the above, it should be emphasized that all summations are written out explicitly and $\alpha$ and $\beta$ are not summed over. Since the graphene Hamiltonian Eq.~\eqref{h0tau} is spin-independent, the trace over spins is rendered only on the electron spin operators $\bm{S}$ in the above equation. We have assumed that the concentration of the magnetic adatoms (\textit{e.g.} from transition metal elements) is dilute enough not to affect the band structure of graphene significantly. In the case of sufficient adatom concentration, graphene electrons could acquire spin-orbit coupling from the adatoms \cite{grapQAHE1,grapQAHE2,ASm1,Asm2} and the calculation here would need to be generalized to account for spin-orbit coupling.

 The explicit summation over spins can be carried out on the electron spin operator $S_{\mu}=(\hbar/2) \sigma_{\mu}$,  by using the trace relation $\rm{Tr}\{\sigma_{\mu}\sigma_{\nu}\}=2\delta_{\mu\nu}$ for Pauli matrices $\sigma_{\mu}$, yielding
\begin{eqnarray}
 \sum_{\mu, \nu ,a,b} (S_{\mu})_{a b}  (S_{\nu})_{b a} \mathbb{S}_{1,\mu} \mathbb{S}_{2,\nu}=\frac{\hbar^2}{2}\mathbb{S}_{1}\cdot\mathbb{S}_{2},
\end{eqnarray}
so that the exchange energy becomes $E_{\alpha \beta}(\boldsymbol{R})=J_{\alpha \beta}(\boldsymbol{R}) \mathbb{S}_{1}\cdot\mathbb{S}_2$, where the exchange coupling strength can be written as (see Appendices~\ref{apexchf0}, \ref{RKKYF0} and  Eq.~\eqref{ImGG1} for details),
\begin{eqnarray}\label{exchJ}
  J_{\alpha \beta}({\bm R})= &&\lambda^2\hbar^2 \int^{\frac{\hbar\Omega}{2}}_{-\frac{\hbar\Omega}{2}}\frac{d\bar{\omega} }{2\pi}\nonumber \\ &&{\rm Im}\left\{{\rm Tr}\left[G^{\rm R}_{\alpha \beta}({\bm R},\bar{\omega})G^<_{ \beta\alpha}(-{\bm R},\bar{\omega})\right]\right\}\;.
 \end{eqnarray}
%

\section{Quasienergy Spectrum and Real-Space Floquet Green's Functions}\label{Calculation of the Green's Funtions}
\begin{figure}
    \begin{center}
            \includegraphics[width=1\columnwidth]{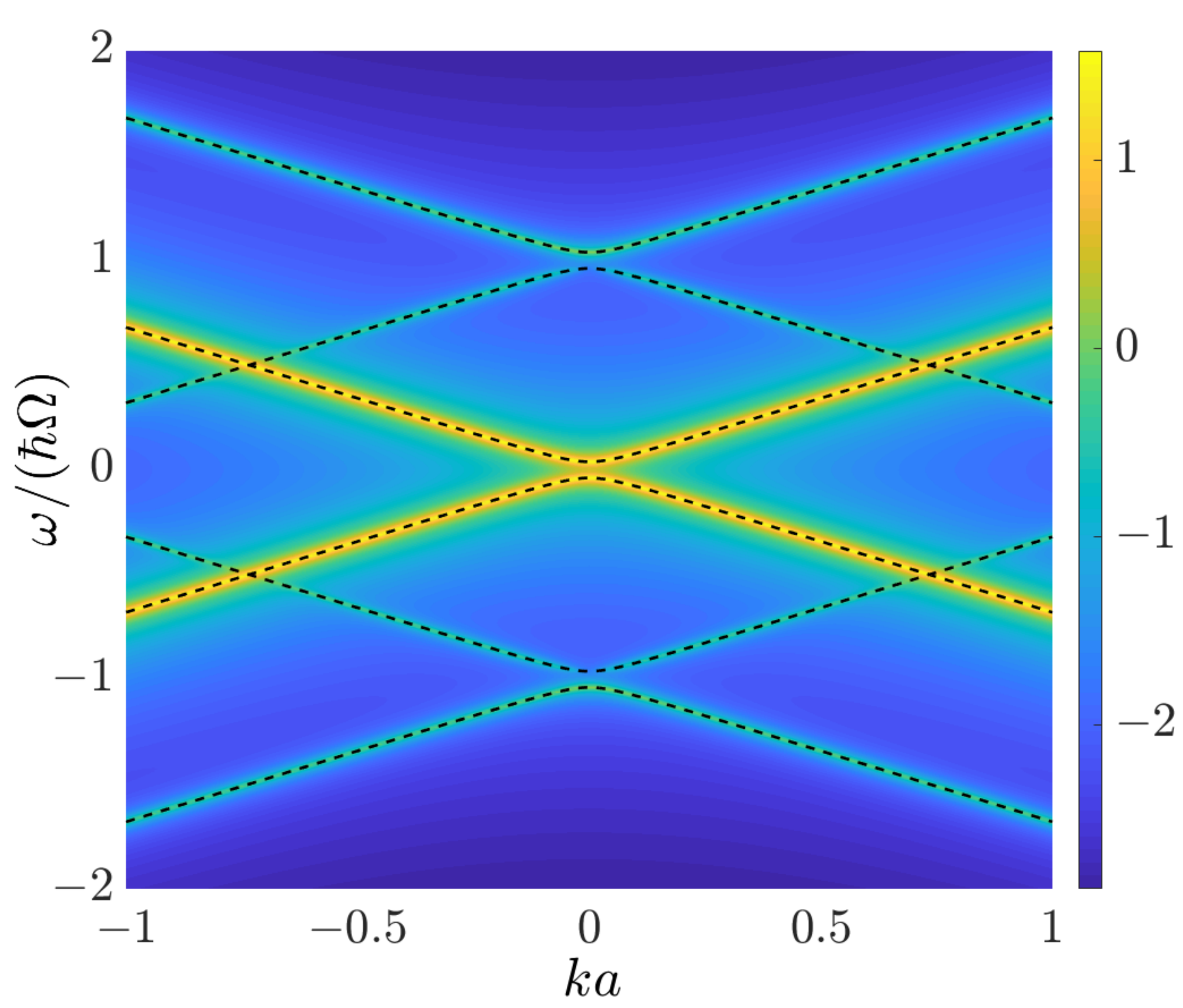}
                \end{center}
                \caption{Time-averaged spectral function $\rho(\bk,\omega)$ [Eq.~\eqref{spectralfunc}] as a function of $\bk$ and $\omega$, obtained within the $F_0$ approximation. The side color bar indicates the magnitude of $\rho(\bk,\omega)$ in logarithmic scale. The dashed lines indicate the Floquet quasienergy dispersions [Eq.~\eqref{qusieng}]. Here the system is driven by a field with strength  $\mathcal{A}=A/\hbar\Omega=0.2$ and frequency  $\hbar\Omega=2.2\gamma$.}
    \label{bands}
\end{figure}
To compute the RKKY coupling from Eq.~\eqref{exchJ}, in this section we derive the real-space Floquet Green's functions of irradiated graphene. The Floquet Green's functions can be obtained directly from the Floquet Hamiltonian. For an arbitrary driving strength, this generally needs to be done numerically and the calculation of Eq.~\eqref{exchJ} would require extensive computational effort, and increasingly so towards stronger driving. In this work, we focus only on the weak drive regime $\mathcal{A} \ll 1$, in which the theory becomes analytically tractable.
To this end, we first project the Floquet Hamiltonian into a new set of basis, which will be called photon-number representation, that makes it more physically transparent. Its main idea can be motivated from the observation that the graphene Floquet Hamiltonian Eq.~\eqref{flm} under CP light breaks into a set of decoupled two-level systems at $\bk = 0$. For instance, at the $K$-point, those two levels comprise the $A$ sublattice component of the $n-1$ Floquet mode and the $B$ sublattice component of the $n$ Floquet mode, they are coupled via $A=v_F e E_0/(\sqrt{2}\Omega)$ but are decoupled from the all other Floquet modes. Hence, we can project the entire Floquet Hamiltonian for all $\bk$ values into the set of basis that diagonalize its block diagonal representation at $\bk=0$. This method was first introduced in Ref.~\cite{busl} and we can express it in the following form consisting of three consecutive unitary transformations:
\begin{eqnarray}\label{us}
\tilde{\mathcal{H}}_{F} =U^{\dagger}_{T,\tau} \mathcal{H}_F U_{T,\tau}\,\,\,\,\,{\rm with}\,\,\,\,\,U_{T,\tau} = U_{1,\tau}U_{2,\tau}U_3,
\end{eqnarray}
where $\tau=\pm1$ and the operators with $\tau=+$ ($\tau=-$) act on the Hamiltonian at the $K$ ($K'$) valley. The definitions of the transformations $U_{1,\tau}, U_{2,\tau}, U_3$ are relegated to Appendix~\ref{unittrans}. The resulting Hamiltonian $\tilde{\mathcal{H}}_{F}$ is in the basis $\cup_{n = -\infty}^{\infty}\{\Phi_{\uparrow,n}, \Phi_{\downarrow,n}\}$, where the $\uparrow, \downarrow$ symbols label a new ``isospin'' degrees of freedom:
\begin{eqnarray}\label{bas1}
\Phi_{\uparrow,n}=i h_+ \phi_{A,n+1}+h_-\phi_{B,n},\nonumber\\
\Phi_{\downarrow,n}=g_+ \phi_{A,n}-i g_-\phi_{B,n-1},
\end{eqnarray}
for the $K$ point and
\begin{eqnarray}\label{bas2}
\Phi_{\uparrow,n}=g_+ \phi_{A,n}-i g_-\phi_{B,n-1},\nonumber\\
\Phi_{\downarrow,n}=-i h_+ \phi_{A,n+1}-h_-\phi_{B,n},
\end{eqnarray}
for the $K'$ point, where
\begin{eqnarray} \label{ghpm}
g_{\pm}=\frac{2A}{\sqrt{4A^2+(\mp\hbar\Omega+\sqrt{4A^2+\hbar^2\Omega^2})^2}},\nonumber\\
h_{\pm}=\frac{\hbar\Omega\mp\sqrt{4A^2+\hbar^2\Omega^2}}{\sqrt{4A^2+(\mp\hbar\Omega+\sqrt{4A^2+\hbar^2\Omega^2})^2}}. \nonumber
\end{eqnarray}
One notices that the explicit matrix forms of $\tilde{\mathcal{H}}_{F}$ for the two valleys happen to be the same when expressed in the bases Eqs.~\eqref{bas1}-\eqref{bas2}.
It is also useful to note that the basis of the $K$ point can be transformed to the basis of the $K'$ point via
\begin{equation}\label{sytrns}
\Sigma_{y}(\dots,\Phi^{K}_{\uparrow,n}, \Phi^{K}_{\downarrow,n},\ldots)^{T}=(\dots,\Phi^{K'}_{\uparrow,n}, \Phi^{K'}_{\downarrow,n},\ldots)^{T}\;,
\end{equation}
where $\Sigma_{y}=i\sigma_{y}\otimes\mathbb{I}_{\infty}$, $\mathbb{I}_{\infty}$ is the identity matrix in the Floquet space, and $\Phi^{K}_{\uparrow,\downarrow}$ and $\Phi^{K'}_{\uparrow,\downarrow}$ correspond to $\Phi_{\uparrow,\downarrow}$ in Eqs.~\eqref{bas1} and \eqref{bas2}, respectively.

Up to this point, there is no approximation and the transformed Hamiltonian $\tilde{\mathcal{H}}_{F}$ captures the same physics as the original Floquet Hamiltonian in Eq.~\eqref{floqh}, but it provides a more  convenient picture of the phenomena at play. The expression of $\tilde{\mathcal{H}}_{F}$ contains the following parameters
\begin{eqnarray}\label{fN}
F_0 &=&\frac{1}{2} \left(\hbar \Omega/\sqrt{4 A^2+\hbar^2 \Omega ^2}+1\right),\nonumber
\\F_1&=&i A/\sqrt{4 A^2+\hbar^2 \Omega ^2},
\\F_2&=&\frac{1}{2} \left(1-\hbar \Omega/\sqrt{4 A^2+\hbar^2 \Omega ^2}\right).\nonumber
\end{eqnarray}
Each of these parameters captures a successive photon process mimicking a tight-binding Hamiltonian in the Floquet-ladder space; {\it i.e.} $F_0$ captures the zero-photon process that opens a gap at $\bk =0$ resulting from time-reversal symmetry breaking (``on-site"). $F_1$ captures the one-photon resonance resulting from the hybridization of the first nearest-neighbor ($|n-m|=1$) Floquet-ladder bands at $\bk=\Omega/(2v_{F})$, while $F_2$ captures the two-photon resonance resulting from the hybridization of the second nearest-neighbor ($|n-m|=2$) Floquet-ladder bands at $\bk=\Omega/v_{F}$. Similarly, there are other coefficients $F_{N>2}$ describing higher-order photon processes, but will not be needed for our present theory.

The values of $F_{N=0,1,2}$ in the  Floquet-ladder tight-binding Hamiltonian $\tilde{\mathcal{H}}_{F}$ depends on the driving strength $\mathcal{A}=A/(\hbar\Omega)$. For consistency with the regime of validity of the continuum Floquet Hamiltonian that sets an upper limit on the value of $\mathcal{A}$ (see Eq.~\eqref{flm} and the relevant discussion in Sec.~\ref{System and the method}),
in the following we will focus on the weak drive regime $\mathcal{A}\ll 1$ which renders $F_{1}$ and $F_{2}$ negligible compared to $F_{0}$ because they are smaller by a factor of $\mathcal{A}$ and $\mathcal{A}^2$.
This approximation is known as the $F_{0}$ approximation~\cite{busl}. Hence, $\tilde{\mathcal{H}}_{F}$ becomes dependent only on $F_{0}$ and takes a block-diagonal form $\tilde{\mathcal{H}}_{F} = \bigoplus_{n = -\infty}^{\infty}\tilde{h}_{n}$, where $\bigoplus$ stands for the matrix direct sum over the Floquet space, with the following $2\times 2$ block Hamiltonian $\tilde{h}_{n}$:
\begin{eqnarray}\label{hf0}
\tilde{h}_{n}=\left[
\begin{array}{cc}
 n \hbar  \Omega+\Delta/2  & e^{i \theta_k } F_0 \hbar v_F k \\
 e^{-i \theta_k }F_0 \hbar v_F k & n \hbar  \Omega -\Delta /2 \\
\end{array}
\right],
\end{eqnarray}
where $\Delta=\sqrt{4 A^2+\hbar^2 \Omega ^2}-\hbar \Omega$ and $\theta_k$ is the angle between $\boldsymbol{k}$ and the $\boldsymbol{K'}-\boldsymbol{K}$ direction. Within the $F_{0}$-approximation the Hamiltonian in Eq.~\eqref{hf0} captures the photon-induced band gap $\Delta$ at the $K$ and $K'$ points and the photon-induced renormalization of the graphene Fermi velocity by $F_{0}$. The quasienergies for $K$ or $K'$ are simply gapped Dirac dispersions shifted by integer multiples of $\hbar\Omega$,
\begin{equation}\label{qusieng}
  \epsilon_{n,\bk}=\pm\sqrt{(F_{0}\hbar v_{F}k)^2+(\Delta/2)^2}+n\hbar\Omega\;.
\end{equation}
The dashed line in Fig.~\ref{bands} shows the quasienergy dispersion for $n = 0$.

Here we can make a connection to the approximate Floquet Hamiltonian $H_{{\rm FM}}$ obtained under the Floquet-Magnus expansion~\cite{FMExpansion} up to second order that is commonly used as starting point for calculation of Floquet dynamics in the high-frequency off-resonant regime. For irradiated Dirac system such as graphene, it gives~\cite{Floqtrans5} a static gapped Dirac Hamiltonian with a gap $\Delta_{{\rm FM}}=A^{2}/(\hbar\Omega)$. On the other hand, the $F_0$ Hamiltonian captures not only the induced gap but also the Fermi velocity renormalization for all Floquet modes $n \in \mathbb{Z}$. For the case of graphene, one can obtain $H_{{\rm FM}}$ from  the $F_{0}$ approximation by expanding Eq.~\eqref{hf0} in $\Delta$ and $F_{0}$ up to second order in $\mathcal{A}$ and then setting $n=0$.

Having obtained the Floquet Hamiltonian, we can now proceed to obtain the real-space Green's functions in the photon-number representation under the $F_0$ approximation. The simple form of the $F_0$ Hamiltonian allows us to obtain analytically tractable Green's functions. We start with the momentum-space Green's functions and  obtain the corresponding real-space counterparts through a Fourier transformation. The retarded Green's function in the momentum space is $\tilde{\mathcal{G}}^{{\rm R}} = \bigoplus_{n = -\infty}^{\infty}\tilde{g}^{{\rm R}}_{n}$, where the $2\times 2$ block Green's function corresponding to the $n^{\mathrm{th}}$ Floquet mode is given by $\tilde{g}^{{\rm R}}_{n}(\boldsymbol{k},\bar{\omega})=[(\bar{\omega}+i \eta)\mathbb{I}_{\sigma}-\tilde{h}_{n}]^{-1}$. In the following, we write the Green's function in terms of the physical, extended-zone frequency ${\omega} = \bar{\omega} -n\hbar\Omega$ with $\tilde{g}^{{\rm R}}(\boldsymbol{k},\omega) \equiv \tilde{g}^{{\rm R}}_{n = 0}(\boldsymbol{k},\omega)$:
\begin{eqnarray}
  &&\tilde{g}^{{\rm R}}(\boldsymbol{k},\omega)=-\frac{1}{D_+(k,\omega)}
\begin{bmatrix}
   \omega+i \eta+\Delta /2 &  \hbar v_F F_0 k e^{i \theta_k} \\
   \hbar v_F F_0 k e^{-i \theta_k}  &   \omega+i \eta-\Delta /2
    \end{bmatrix}, \nonumber \\ \label{Greenk}
\end{eqnarray}
where
$D_{\pm}(k,\omega)=(\Delta /2)^2+(\hbar v_F F_0 k)^2-(\omega \pm i\eta)^2$.
 \begin{figure}
\centering
            \includegraphics[width=1\columnwidth]{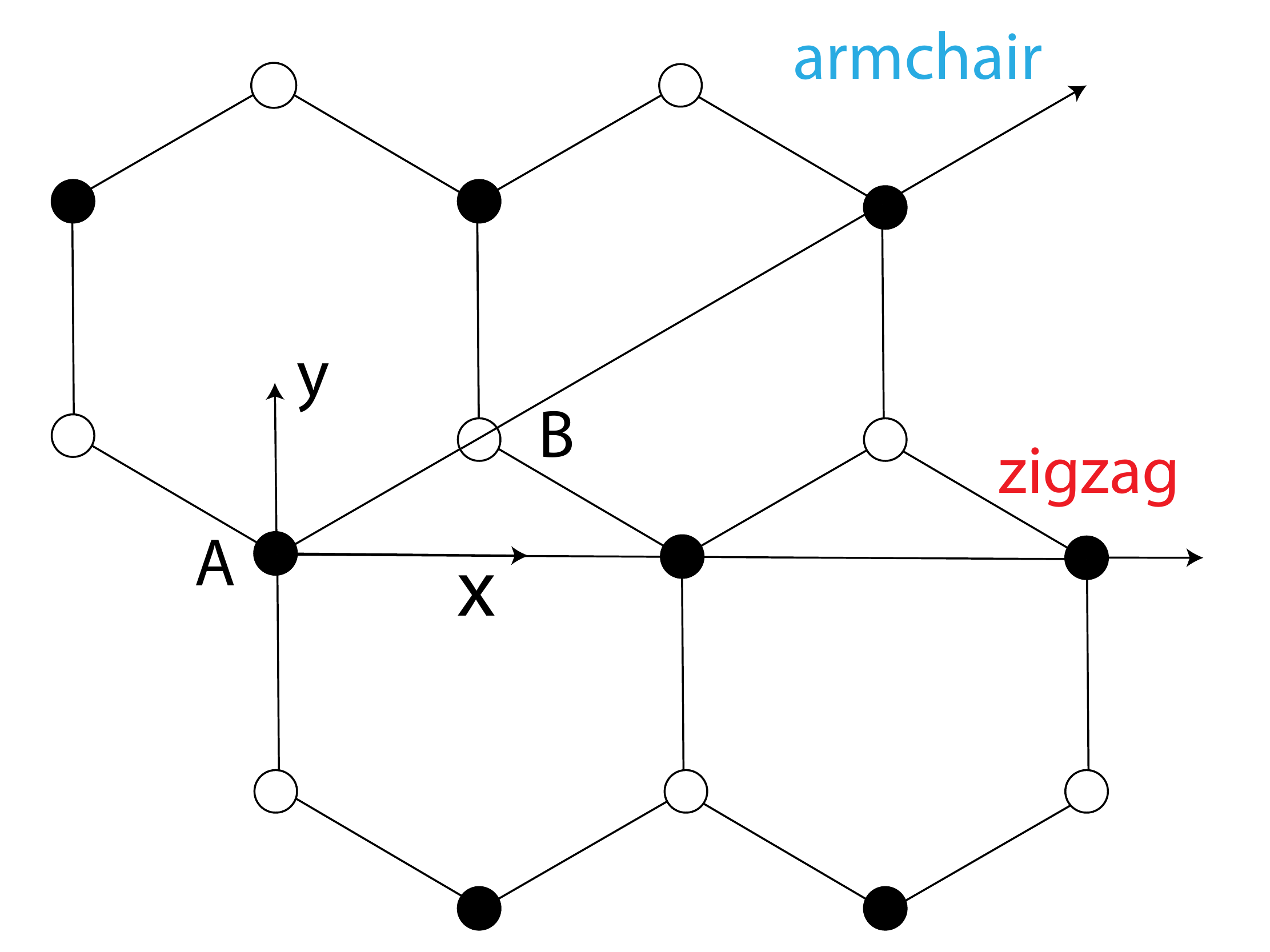}
    \caption{Honeycomb lattice structure of graphene. The filled circles represent the atoms of sublattice $A$ while the unfilled ones represent the atoms of sublattice $B$. The zigzag and armchair directions in the honeycomb structure are indicated with arrows.}
\label{fig1}
\end{figure}
The electronic spectral weight in driven systems can be characterized by the time-averaged spectral function, which can be obtained from the Floquet Green's function in the original basis as
%
\begin{eqnarray}\label{spectralfunc}
\rho(\bk,\omega) = -\frac{1}{\pi}\sum_{\alpha \in \{A,B\}}{\rm Im}\left\{\left[\mathcal{G}^{{\rm R}}_{\alpha\alpha}(\bk,{\omega})\right]_{00}\right\}\;.
\end{eqnarray}
%
The color intensity in Fig.~\ref{bands} shows $\rho(\bk,\omega)$ for valley $K$ obtained with $\mathcal{G}^{{\rm R}}(\bk,\bar{\omega}) = U_{T,+}\tilde{\mathcal{G}}^{{\rm R}}(\bk,\bar{\omega})U^{\dag}_{T,+}$ in the $F_0$ approximation.


The real-space Green's functions for each valley are obtained by their corresponding Fourier transformation of the momentum-space Green's functions
\begin{eqnarray}\label{all}
\tilde{g}(\boldsymbol{R},\omega) &=&\int \frac{d\boldsymbol{k}}{(2\pi)^2} e^{i \boldsymbol{k}\cdot \boldsymbol{R}} \tilde{g}(\boldsymbol{k},\omega).
\end{eqnarray}
Substituting Eq.~\eqref{Greenk} into the above, we obtain the retarded Green's function for each Floquet mode in the photon-number representation (see Appendix~\ref{rsgf} for details of the derivation),
 \begin{eqnarray}\label{pm1}
&&[\tilde{g}^{\rm R}({\bm R},\omega)]_{\uparrow\uparrow ,\downarrow\downarrow}=-\zeta[\omega+i\eta\pm\Delta/2]\chi_0(R,\omega),
\end{eqnarray}
\begin{equation}\label{pm2}
[\tilde{g}^{\rm R}({\bm R},\omega)]_{\uparrow\downarrow,\downarrow\uparrow}
=-\zeta e^{\pm i\theta_R}\chi_1(R,\omega),
\end{equation}
where $+$ ($-$) on the right-hand side of Eqs.~\eqref{pm1}-\eqref{pm2} corresponds to $\uparrow\uparrow$ ($\downarrow\downarrow$) and $\uparrow\downarrow$ ($\downarrow\uparrow$) configurations of the isospins in the left-hand side of these equations [see Eqs.~\eqref{bas1} and \eqref{bas2} for isospin definition], $\zeta^{-1}=2\pi(\hbar v_F F_0)^2$, $\theta_R$ is the angle between ${\bm R}$ and the $x$-axis, and
\begin{eqnarray}
&&\chi_0(R,\omega)=\frac{\pi i}{2}{H}^{(1)}_0\left[\kappa(\omega)R\right],
\\ &&\chi_1(R,\omega)=-\frac{\pi}{2}  \hbar v_F F_0\kappa(\omega){H}_1^{(1)}\left[\kappa(\omega) R\right],\\
&&\kappa(\omega)=\frac{{\rm sgn}(\omega)}{\hbar v_F F_0}\sqrt{(\omega+ i\eta)^2-\frac{\Delta^2}{4}},\label{pm3}
\end{eqnarray}
where ${H}^{(1)}_0$ and ${H}^{(1)}_1$ are the zeroth-order and first-order Hankel functions of the first kind, respectively.

We assume thermalization of  the irradiated system is achieved in NESS through a coupling to an external fermion bath  (\textit{e.g.}, a metallic lead). The Floquet-Keldysh formalism enables one to determine the Floquet lesser Green's function in the momentum space as
\begin{eqnarray}
\mathcal{G}^{<}(\boldsymbol{k},\bar{\omega})&=&\mathcal{G}^{\rm R}(\boldsymbol{k}, \bar{\omega}) {\Sigma}^{<}(\bar{\omega})\mathcal{G}^{\rm A}(\boldsymbol{k}, \bar{\omega}),
\end{eqnarray}
where ${\Sigma}^{<} = -{\Sigma}^{\mathrm{R}}F+F {\Sigma}^{\mathrm{A}}$ is the lesser self-energy, and $F(\bar{\omega}) = \bigoplus_{n = -\infty}^{\infty} f(\bar{\omega}-n\hbar\Omega)\mathbb{I}_{\sigma}$ is the Floquet representation of the equilibrium distribution with $f(\omega)=1/[e^{(\omega-\mu)/(k_B T)}+1]$. In the wide-band approximation for the fermion bath, the retarded self-energy takes the momentum-independent form ${\Sigma}^{\mathrm{R}} = -i\eta \mathbb{I}_{\sigma}\otimes\mathbb{I}_{\infty}$, and the lesser self-energy becomes  ${\Sigma}^{<} = 2i\eta \bigoplus_{n = -\infty}^{\infty} f(\bar{\omega}-n\hbar\Omega)\mathbb{I}_{\sigma}$.

Transforming to the photon-number representation, we find the lesser Green's function 
\begin{eqnarray} \label{glessF0}
&\tilde{g}^{<}(\bk,\omega)=2i\eta F_0 f(\omega) \tilde{g}^{\rm R}(\boldsymbol{k}, \omega) \tilde{g}^{\rm A}(\boldsymbol{k}, \omega).
\end{eqnarray}

The real-space lesser Green's function is then obtained from the Fourier transformation of Eq.~\eqref{glessF0} (see Appendix~\ref{rsgf}), yielding
\begin{eqnarray}\label{gl1}
  &&[\tilde{g}^<({\bm R},\omega)]_{\uparrow\uparrow ,\downarrow\downarrow}= F_0 f(\omega)\zeta\left\{\frac{(\omega\pm\Delta /2)^2+\eta^2}{2\omega}\nonumber\right.\\
&&\left.\times[\chi_0( R,\omega)-\chi_0(R,-\omega)]+i\eta\{[1-B(\omega)]\chi_0(R,\omega)\right.\nonumber
\\
&&\left.+[1+B(\omega)]\chi_0(R,-\omega)\}\right\},
\end{eqnarray}
\begin{eqnarray}\label{gl2}
[\tilde{g}^<({\bm R},\omega)]_{\uparrow\downarrow,\downarrow\uparrow}=F_0 f(\omega)e^{\pm i \theta_R}\zeta\left[\chi_1(R,\omega)-\chi_1(R,-\omega)\right],\nonumber\\
\end{eqnarray}
where $B(\omega)=(\Delta^2/2-2\omega^2+2\eta^2)/(4i\eta\omega)$.

With all the required Green's functions obtained, we return to the expression of the RKKY coupling in Eq.~\eqref{exchJ}. The main contribution to the RKKY coupling in graphene
comes from the $K$ and $K'$ points. Therefore the total real-space Green's function consists of contributions from both valleys,
\begin{eqnarray}\label{FourG}
G({\bm R},\bar{\omega}) &=&\int \frac{d\bk}{(2\pi)^2}  e^{i \bk \cdot {\bm R}} \left[e^{i {\bm K}\cdot {\bm R}}\mathcal{G}_{+}({\bm k},\bar{\omega})\right.\nonumber\\
&&\left. + e^{i {\bm K'}\cdot {\bm R}}\mathcal{G}_{-}(\bk,\bar{\omega})\right],
 \end{eqnarray}
 where $\mathcal{G}_{+}$ ($\mathcal{G}_{-}$) denotes the Green's function associated with the ${\bm K}$ (${\bm K'}$) Dirac points. The above Green's function $G({\bm R},\omega)$, which enters Eq.~\eqref{exchJ}, is written in the original Floquet representation. A crucial observation is that the  unitary transformation $U_{T,\pm}$ in Eq.~\eqref{us} for the Hamiltonian is $\bm{k}$-independent; this allows the same transformation to carry over to the transformation for real-space Green's function also. Through this transformation we can formally express all the Green's functions in Eq.~\eqref{exchJ} from the original Floquet representation into the photon-number representation. We choose the basis for the ${\bm K'}$ point, Eq.~\eqref{bas2}, as the common basis. Then, applying the transformation $U_{T,-}$ [Eq.~\eqref{us}] and defining
$\tilde{G}({\bm R},\bar{\omega})  = U^{\dag}_{T,-}G({\bm R},\bar{\omega})U_{T,-}$, Eq.~\eqref{FourG} becomes
\begin{eqnarray}\label{FourG2}
\tilde{G}({\bm R},\bar{\omega}) &=&\int \frac{d\bk}{(2\pi)^2}  e^{i \bk \cdot {\bm R}} \left[e^{i {\bm K}\cdot {\bm R}}\Sigma_{y}\tilde{\mathcal{G}}({\bm k},\bar{\omega})\Sigma^{\dag}_{y}\right.\nonumber\\
&&\left. + e^{i {\bm K'}\cdot {\bm R}}\tilde{\mathcal{G}}(\bk,\bar{\omega})\right],
 \end{eqnarray}
 where  we have used Eq.~\eqref{sytrns}
and have written  $\tilde{\mathcal{G}}(\bk,\bar{\omega}) \equiv \tilde{\mathcal{G}}_+(\bk,\bar{\omega})=\tilde{\mathcal{G}}_-(\bk,\bar{\omega})$, since $\tilde{\mathcal{G}}_{\pm}(\bk,\bar{\omega})$ have the same explicit matrix form in the photon-number representation.
With Eq.~\eqref{FourG2} and
the $F_{0}$ approximation, Eq.~\eqref{exchJ} takes on the simplified form
 (see Appendix~\ref{RKKYF0} for details),
\begin{eqnarray}\label{exchJ2}
  J_{\alpha \beta}({\bm R})= &&\lambda^2\hbar^2 F^2_{0} \int^{\frac{\hbar\Omega}{2}}_{-\frac{\hbar\Omega}{2}}\frac{d\bar{\omega}}{2\pi} \\ &&{\rm Im}\left\{{\rm Tr}\left[\tilde{G}_{\alpha\beta}^{\rm R}({\bm R},\bar{\omega})\tilde{G}_{\beta\alpha}^<(-{\bm R},\bar{\omega})\right]\right\}\;.\nonumber
 \end{eqnarray}
 Under the $F_{0}$ approximation, the Green's function in Eq.~\eqref{exchJ2} are block diagonal in the Floquet index $n$ consisting of the $2\times2$ Green's functions $\tilde{g}_{n}({\bm R, \bar{\omega}})$.
 This allows us to further express Eq.\eqref{exchJ2} in terms of the extended zone frequency $\omega\in (-\infty,\infty)$, 
\begin{eqnarray}\label{exch2}
  J_{\alpha \beta}({\bm R})=&& \lambda^2\hbar^2 F^2_{0} \int^{\infty}_{-\infty}\frac{d\omega }{2\pi}\\ &&{\rm Im}\left\{[\tilde{g}_{F_0}^{\rm R}({\bm R},\omega)]_{\alpha\beta}[\tilde{g}_{F_0}^<(-{\bm R},\omega)]_{\beta\alpha}\right\}\;,\nonumber
 \end{eqnarray}
where
\begin{equation}\label{FourG3}
\tilde{g}_{F_0}({\bm R},\omega)=e^{i {\bm K}\cdot {\bm R}}\sigma_{y}\tilde{g}({\bm R},\omega)\sigma_{y}+ e^{i {\bm K'}\cdot {\bm R}}\tilde{g}(\bm R,\omega),
 \end{equation}
 with $\tilde{g}({\bm R},\omega)$ given by Eq.~\eqref{all}.
 In the above, we have used a common numerical representation $\{+1,-1\}$ for the pseudospins $\{A,B \}$ and the $K'$ isospins $\{\uparrow,\downarrow\}$.
In the absence of irradiation, Eq.~\eqref{exch2} can be reduced to its equilibrium limit \cite{satpathy1,serami} as shown in Appendix~\ref{RKKYF0}.

\section{Numerical and Analytical Results}\label{Numerical results}
\begin{figure*}

   \begin{center}
            \includegraphics[width=\textwidth]{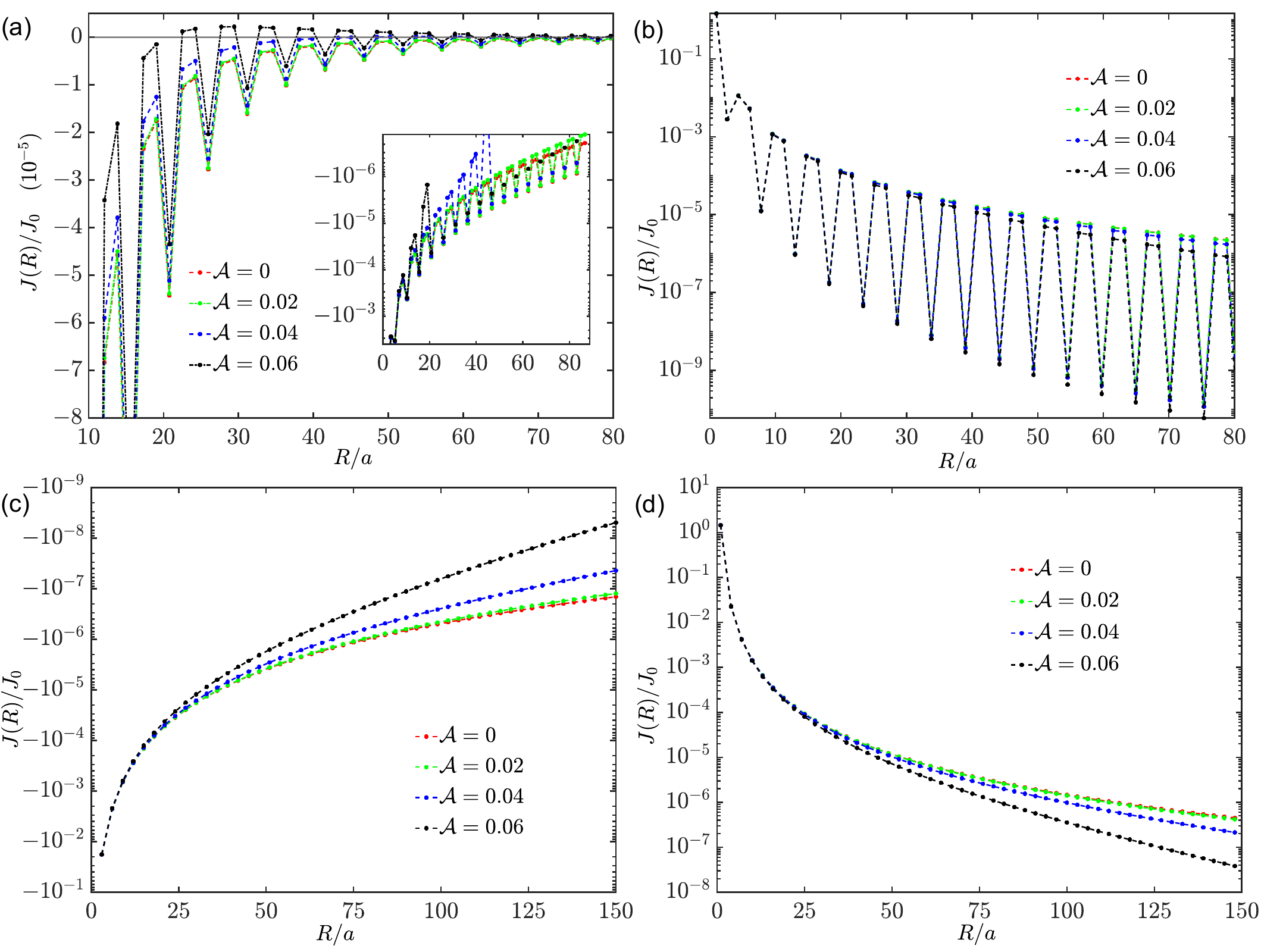}
                \end{center}
                \caption{Time-averaged RKKY coupling $J(R)/J_{0}$ for undoped graphene under irradiation. The driving field is taken to have a frequency $\hbar\Omega=6.6\gamma$ and strength $\mathcal{A} \in [0,0.06]$, with $\mathcal{A} = 0$ being the equilibrium case. The panels correspond to cases with the impurities located at different sublattice sites ($A$ or $B$) and separated along different directions (zigzag or armchair): (a) $AA$ zigzag; (b) $AB$ zigzag; (c) $AA$ armchair; (d) $AB$ armchair. The inset in (a) shows the negative values of $J(R)/J_{0}$ over the full range of $R$ down to the smallest separation.
$J_{0}$ is given in Eq.~\eqref{alpha}.}
\label{fig2}

\end{figure*}

Eq.~\eqref{exch2} allows for efficient numerical and analytical evaluation of the non-equilibrium RKKY coupling. In this section we present numerical results for $J(\bm R)$ and perform additional  approximations to obtain analytic results. We consider both undoped and doped graphene for different values of $\mathcal{A}$, for impurities located at the same sublattice sites or at different sublattice sites, with the separation $\bm R$ along the zigzag or armchair direction. In all cases we also include results for the undriven case with $\mathcal{A} = 0$, which are consistent with the equilibrium results  \cite{serami,blackshaffer,satpathy1,satpathy2} in the literature.

\subsection{Undoped Graphene, $E_{F}=0$}\label{intrinsic}

The panels of Fig.~\ref{fig2} show our numerically evaluated RKKY coupling $J$ for different cases of impurity locations and separation directions for Fermi level located at the Dirac point. The  choice of the range of driving strength $\mathcal{A}$ is informed by its upper limit   $\mathcal{A} \ll 3\gamma/\hbar\Omega < 1$ (see Sec.~\ref{System and the method})
for the particular frequency value $\hbar\Omega =6.6 \gamma$ we have chosen. For this large frequency value the driving field is off-resonant and the system is in the weak drive regime.

When the magnetic impurities are located on the same sublattice ($AA$ or $BB$) and connected through the zigzag direction [Fig.~\ref{fig2}(a)],
$J$ displays short-scale oscillations as a function of the impurity separation $R$ for all the driving strengths $\mathcal{A}$ studied. This is a general feature in systems with a multi-valley band structure \cite{kittelbook,RKKYvalleys1} and arise from intervalley scattering. Graphene's band structure contains two valleys $K$ and $K'$ that originate from the honeycomb lattice structure. As such, these short-scale oscillations persist with and without irradiation and have an intrinsic period that is independent of Fermi level.
They have a period $2\pi/|{\bm K}-{\bm K'}|=3\sqrt{3}a/2$, where $|{\bm K}-{\bm K'}|$ is the distance between two adjacent valleys in the Brillouin zone and $a$ is given in Eq.~\eqref{h0tau}. It is convenient to take the $k_{x}$-axis along an adjacent pair of $K$ and $K'$ with the origin chose at the $M$-point between them, such that ${\bm K}-{\bm K'}=2{\bm K}$.
One may notice that these short-range oscillations can only manifest for certain choices of impurity separation in graphene, namely, when the phase difference $({\bm K}-{\bm K'})\cdot \Delta{\bm R} \ne 2\pi$, where $\Delta{\bm R}$ is the increment that gives the next point of impurity separation on the lattice, and is given by the distance between two adjacent impurity sites along the zigzag or armchair direction. Along one of these directions, it is the vector joining adjacent $A$ sublattice sites for impurity spins on the same sublattice, and the vector joining adjacent $A$ and $B$ sublattice sites for impurity spins on different sublattices.
As a result, these sharp oscillations  appear in the case of the $AA$ ($BB$) impurities separated along the zigzag direction, which has $({\bm K}-{\bm K'})\cdot \Delta{\bm R}=2\pi/3$, in addition to the overall decrease with increasing $R$.
For weak driving strengths $\mathcal{A}$, we observe that irradiation suppresses the RKKY  coupling for all values of $R$ while maintaining its sign, hence remaining ferromagnetic in character $J(\bm R)<0$ like in equilibrium. Interestingly however, we find that for large enough driving $J(\bm R)$ can flip sign and display antiferromagnetic character within a rather large
range of $R$, as seen for $\mathcal{A} = 0.06$ in Fig.~\ref{fig2}(a).

To gain further insights into this behavior, we proceed to derive approximate analytic results. Our results in the following are valid up  to $\mathcal{O}(\Delta^2)$ in the weak drive $\mathcal{A} \ll 1$ regime when the gap  $\Delta/\hbar\Omega \simeq \mathcal{A}^2 \ll 1$.
Taking $\eta\rightarrow 0$ in Eq.~\eqref{exch2} with the expressions of the Green's functions $[\tilde{g}^{\rm R}({\bm R},\omega)]_{\uparrow\uparrow,\downarrow\downarrow}$ and $[\tilde{g}^{<}({\bm R},\omega)]_{\uparrow\uparrow,\downarrow\downarrow}$ from Eqs.~\eqref{pm1} and \eqref{gl1}, and defining $z=-\alpha_{\Delta}\sqrt{|(2\omega/\Delta)^2- 1|}$, the RKKY coupling for  $\alpha=\beta=A$ can be written as $J({\bm R})=\hat{J}_{0}({\bm R})+\bar{J}_{0}({\bm R})$ with
\begin{eqnarray}\label{ani1}
&&\hat{J}_{0}({\bm R})=-\frac{4}{\pi}J_{+}F_0^2 \left(\frac{a}{R}\right)^3\int_{0}^{\alpha_{\Delta}}\BK(z)\BI(z)z\sqrt{z^2+\alpha_{\Delta}^2}dz\nonumber\\
&&-\frac{4}{\pi}J_{-} \left(\frac{ a}{2 l_{\Delta} }\right)^2\left(\frac{a}{R}\right)\int_{0}^{\alpha_{\Delta}}\BK(z)\BI(z)z(z^2+\alpha_{\Delta}^2)^{-\frac{1}{2}}dz\;,\nonumber\\
\end{eqnarray}
and
\begin{eqnarray}\label{Jpm1}
&&\bar{J}_{0}({\bm R})=-2J_+F_0^2\left(\frac{a}{R}\right)^3\int^{\infty}_{\alpha}\BY(z)\BJ(z)z\sqrt{z^2+\alpha_{\Delta}^2}dz\nonumber\\
&&-2J_- \left(\frac{ a}{2 l_{\Delta} }\right)^2\left(\frac{a}{R}\right)\int^{\infty}_{\alpha}\BY(z)\BJ(z)z(z^2+\alpha_{\Delta}^2)^{-\frac{1}{2}}dz.\nonumber\\
\end{eqnarray}
In Eqs.~\eqref{ani1}-\eqref{Jpm1}, $\BJ,\BY$ are the zeroth-order Bessel and Neumann functions, $\BI, \BK$ are the zeroth-order modified Bessel functions of the first and second kind.
We have also made use of ${\bm K}-{\bm K'}=2{\bm K}$ and defined $J_\pm=J_0[1\pm\cos{(2\boldsymbol{K}\cdot {\bm R})}]$ with
\begin{eqnarray}\label{alpha}
&&J_0=\frac{a}{2\pi\hbar v_{F}}\left(\frac{\lambda\hbar}{4a^2}\right)^2,\,\,\,\,\,l_{\Delta}=\frac{\hbar v_{F}}{\Delta}\,\,\,\,{\rm and}\,\,\,\,\,\alpha_{\Delta}=\frac{R}{2F_0 l_{\Delta}}\;.\nonumber\\
\end{eqnarray}
Here $\l_{\Delta}$ is a characteristic length introduced by irradiation and is inversely proportional to the driving strength $\mathcal{A}$.
In Eq.~\eqref{ani1}, since  $z<\alpha_{\Delta}\ll 1$, we can use the small argument expansions of the $\BK$ and $\BI$ and expand the integrands up to second order in $z$. After integration we find both integrals are $\sim (\Delta/2)^3$ and are negligible, hence $J({\bm R})\approx \bar{J}_{0}({\bm R})$ . In Eq.~\eqref{Jpm1} for $\bar{J}_0(\bm R)$, since $\Delta/\hbar \Omega$ is small, $\alpha_{\Delta}\ll 1$ will be satisfied for a large range of $R$. We can therefore extend the lower integration limit from $\alpha$ to $0$, which incur an error $\sim (\Delta/2)^3$ that is negligible. Now $J(\bm R)$ can be written as
\begin{eqnarray}
&&J({\bm R})=-2J_+ F_0^2 \left(\frac{a}{R}\right)^3\int^{\infty}_{0}\BY(z)\BJ(z)z^2dz\\
&&-(J_++2J_-) \left(\frac{ a}{2l_{\Delta}}\right)^2 \left(\frac{a}{R}\right) \int^{\infty}_{0}\BY(z)\BJ(z)dz,\nonumber
\end{eqnarray}
which is analytically integrable leading to
\begin{eqnarray}\label{exa1}
\frac{{J}({\bm R})}{J_{0}}=&-&\frac{F_0^2}{8} \left(\frac{a}{R}\right)^3\left[1+\cos{(2{\bm K}\cdot {\bm R})}\right]\\
&+&\frac{1}{2}\left(\frac{ a}{2 l_{\Delta}}\right)^2\left(\frac{a}{R}\right) \left[3-\cos{(2{\bm K}\cdot {\bm R})}\right].\nonumber
\end{eqnarray}
When $\Delta=0$ and $F_{0}=1$, Eq.~\eqref{exa1} recovers the
the equilibrium RKKY coupling, with $J({\bm R})$ decaying as $1/R^{3}$ modulated by the short-range oscillation factor $[1+\cos{(2{\bm K}\cdot {\bm R})}]$ that arises from intervalley scattering. Notice that these oscillations do not cause any sign change in $J$ as a function of $R$ and  the equilibrium RKKY coupling retains its ferromagnetic character for all $R$~\cite{satpathy1,serami,blackshaffer}. Irradiation modifies this behavior in two interesting ways. In Eq.~\eqref{exa1},   the first term, which is inherited from the  equilibrium contribution, is renormalized by  $F^{2}_{0}$ which decreases with increasing driving strength $\mathcal{A}$.
The second term is a new contribution due to irradiation. It decays more slowly as $1/R$ and increases as $\Delta^2$ with the driving strength $\mathcal{A}$. It displays the same oscillation period since this aspect is uniquely determined by the graphene lattice and the impurity orientation. The interplay between these two terms in Eq.~\eqref{exa1} explains the numerically observed RKKY coupling in Fig.~\ref{fig2}(a) as follows. We notice that the second term has an opposite sign and partly cancels the first, ferromagnetic-like term. Increasing $\mathcal{A}$ thus suppresses the RKKY coupling, as we have found numerically. If $\mathcal{A}$ is large enough, the second term becomes more dominant and the RKKY coupling switches to an antiferromagnetic character. The persistence of this behavior for large $R$ values as observed in Fig.~\ref{fig2}(a) is due to the much slower $1/R$ dependence.

We now consider the case for two impurities located at opposite sublattice sites ($AB$ or $BA$) connected through the zigzag direction. Fig.~\ref{fig2}(b) shows the numerically calculated $J({\bm R})$ for the $AB$ zigzag case, which displays short-range oscillations  similar to the $AA$ zigzag case since $({\bm K}-{\bm K'})\cdot \Delta{\bm R} \neq 2\pi$ also for this configuration. As opposed to the $AA$ zigzag case, the exchange coupling for the $AB$ case is antiferromagnetic in character in equilibrium, and remains so under weak driving $\mathcal{A} \ll 1$. Following similar steps as in the $AA$ zigzag case, we obtain the following analytical result for the $AB$ zigzag RKKY coupling
\begin{eqnarray}\label{exa2}
\frac{{J}({\bm R})}{J_{0}}=&&[1-\cos{(2{\bm K}\cdot {\bm R})}]\nonumber\\&&\times \left\{\frac{3F_0^2}{8}\left(\frac{a}{R}\right)^3-\frac{1}{2}\left(\frac{ a}{2 l_{\Delta}}\right)^2 \left(\frac{a}{R}\right)\right\}.
\end{eqnarray}
The above recovers the equilibrium case~\cite{satpathy1,serami,blackshaffer} in the absence of irradiation ($\Delta=0$ and  $F_{0}=1$), showing an antiferromagnetic coupling. Similar to the $AA$ case, Eq.~\eqref{exa2} consists of two terms, with the second term $\sim 1/R$ arising purely from irradiation. It is opposite in sign and increases as $\Delta^2$ with the driving strength, causing a suppression of  $J({\bm R})$ with increasing $\mathcal{A}$. For weak driving in the range of $\mathcal{A}$ considered, this second term does not become large enough to dominate the first term for all values of $\bm R$. As a result, $J({\bm R})$ does not change sign as observed in Fig.~\ref{fig2}(b), and the irradiated case resembles qualitatively the equilibrium case.

The effect of the relative orientation between the two impurities on the graphene lattice is brought out in the case when the separation is along the armchair direction. Figs.~\ref{fig2}(c)-(d) show the cases when the impurities are located at the same sublattice sites $AA$ and different sublattice sites $AB$. The separation $\Delta{\bm R}$ between the two impurities satisfy $({\bm K}-{\bm K'})\cdot \Delta{\bm R} =2\pi$, and as a result $J({\bm R})$ does not display any short-scale oscillations. Its smooth decay profile resembles the RKKY coupling in conventional insulators~\cite{ins,ins1,ins2,ins3}.
The RKKY coupling remains ferromagnetic for the $AA$ case and antiferromagnetic for the $AB$ case in the weak drive regime, with magnitudes that decrease with increasing $\mathcal{A}$.
We also obtain the following analytic results for the $AA$ case
\begin{equation}\label{exa3}
\frac{{J}({\bm R})}{J_{0}}=-\frac{F_0^2}{4}\left(\frac{a}{R}\right)^3+\left(\frac{ a}{2l_{\Delta} }\right)^2\left(\frac{a}{R}\right),
\end{equation}
and the $AB$ case
\begin{eqnarray}\label{exa4}
\frac{{J}({\bm R})}{J_{0}}=\frac{3 F_0^2}{4}\left(\frac{a}{R}\right)^3-\left(\frac{a}{2l_{\Delta} }\right)^2\left(\frac{a}{R}\right) .
\end{eqnarray}
Except for the absence of the short-range oscillation factors, the above expressions for the armchair case resemble those in the zigzag cases, with similar dependence on $R$, $F_0$ and $\Delta$ in each of the two constituent terms. Eqs.~\eqref{exa1}-\eqref{exa4} are obtained under the assumption $\alpha_{\Delta} \ll 1$. For a given driving strength $\mathcal{A}$, this means that our analytic results are expected to hold for $R \ll 2F_{0} l_{\Delta}$. Comparing with Fig.~\ref{fig2}, we find that Eqs.~\eqref{exa1}-\eqref{exa4} show a good agreement with the numerical results up to  $R \lesssim F_{0} l_{\Delta}$ for $\mathcal{A} \leq 0.04$. For larger $R$, the analytic results Eqs.~\eqref{exa1}-\eqref{exa4} start to deviate from the numerical results. When $R$ is large such that $\alpha_{\Delta} \gtrsim 1$, our numerical  results of $J$ exhibit an exponential decay with $R$
  as $e^{-2\alpha_{\Delta}}$ (see Appendix~\ref{rsgf}), similar to the case of undoped graphene with a gap in equilibrium~\cite{GFR1,gapexch2}. The absence of such an exponential factor in Eqs.~\eqref{exa1}-\eqref{exa4} is due to $e^{-2\alpha_{\Delta}}\approx 1$ in the regime $\alpha_{\Delta} \ll 1$.

To summarize the above analyses of different impurity configurations on undoped graphene, we have found that
the main effect of irradiation under weak driving $\mathcal{A} \ll 1$ is to decrease the magnitude of the RKKY coupling from its equilibrium value. Interestingly, for the particular $AA$ zigzag case, we find that a moderately strong drive (still within $\mathcal{A} \ll 1$) can result in a sign change of the RKKY interaction. In the next section, we will present our results for the doped case with a finite Fermi energy.
\subsection{Doped Graphene, $E_{F}\ne 0$}\label{doped}
\begin{figure*}
 \begin{center}
            \includegraphics[width=\textwidth]{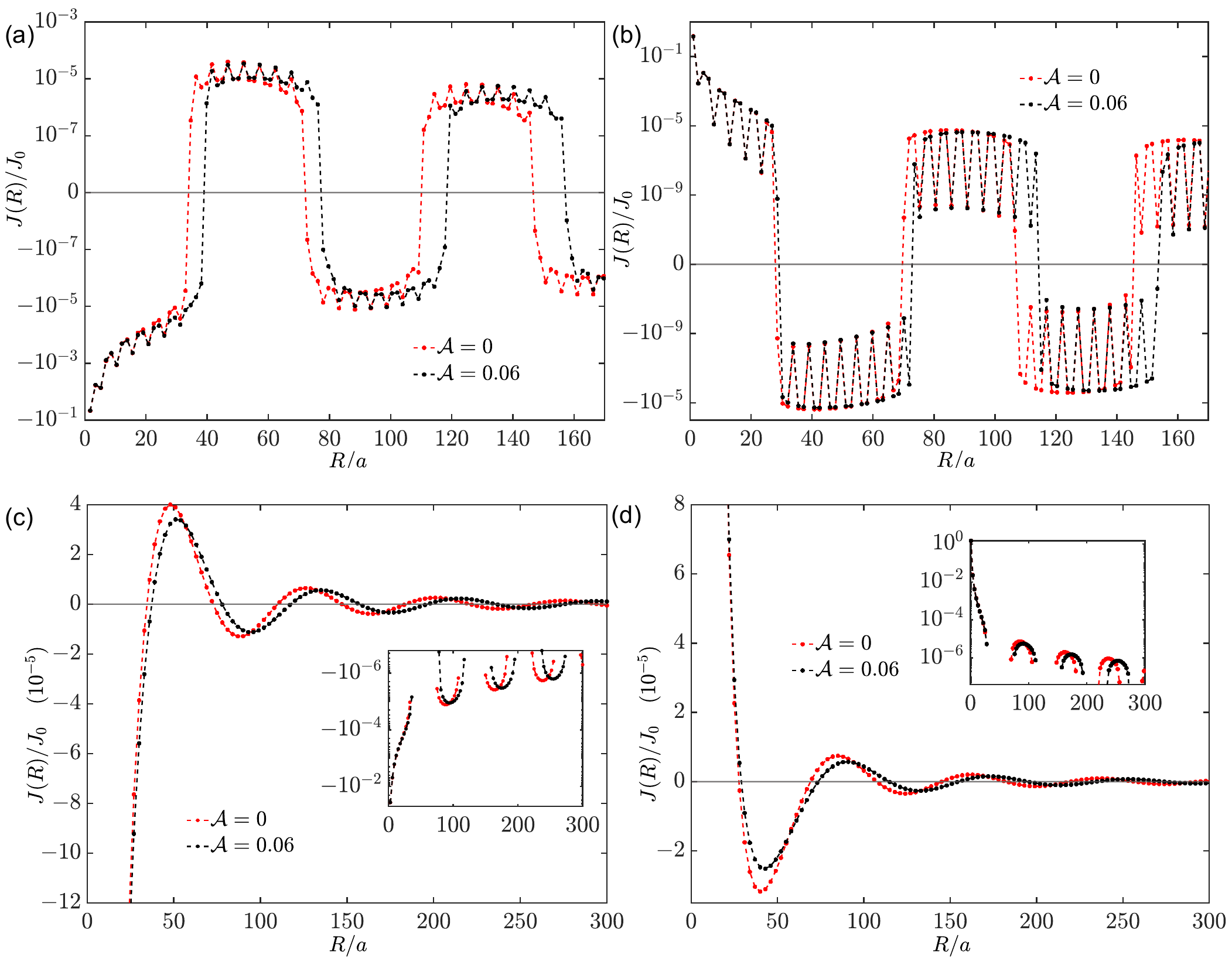}
                \end{center}
                \caption{Time-averaged RKKY coupling $J(R) /J_{0}$ for doped graphene under irradiation with a Fermi energy $E_{F}=0.065\gamma$. The driving field is taken to have a frequency $\hbar\Omega=6.6\gamma$ and strength $\mathcal{A}  = 0$ and $0.06$. For the latter value of $\mathcal{A}$ the photon-induced gap $\Delta=0.71E_{F}$. The panels correspond to cases with the impurities located at different sublattice sites ($A$ or $B$) and separated along different directions (zigzag or armchair): (a) $AA$ zigzag; (b) $AB$ zigzag; (c) $AA$ armchair; (d) $AB$ armchair.
The insets in (c) and (d) show the negative and positive values of $J(R)/J_{0}$ respectively over the full range of $R$ down to the smallest separation. }
\label{fig3}
\end{figure*}

In our calculations we take the Fermi level to be fixed by coupling to the fermion bath  (\textit{e.g.} through attachment to a metallic lead). In addition to the short-scale oscillations resulting from intervalley scattering for the zigzag case, the RKKY interaction displays an additional, longer-range oscillation arising from the intravalley scattering processes near the Fermi surface of each valley. This long-range RKKY oscillation is familiar to the usual case as it fluctuates between both signs as a function of $R$. Figs.~\ref{fig3}(a)-(b) show the $AA$ and $AB$ zigzag cases respectively with a Fermi level  above the photon-induced band gap, $E_{F}>\Delta/2$. The longer-range oscillation is due to the $q = 2k_F$ intravalley scattering and has a period given by $\pi/k_{F}$, where $k_{F}$ is the Fermi wave vector. While the short-range oscillation period remains unchanged under irradiation, we find that the longer-range oscillation period increases with $\mathcal{A}$. This is due to the increased photon-induced gap $\Delta$ with $\mathcal{A}$, resulting in a smaller Fermi surface and a smaller $k_{F}=\sqrt{E^2_{F}-(\Delta/2)^2}/(\hbar v_{F})$ when $E_F$ is fixed. Figs.~\ref{fig3}(c)-(d) show the RKKY coupling respectively for impurities at $AA$ and $AB$ along the armchair direction with a Fermi energy $E_{F}>\Delta/2$. The oscillation is marked by the absence of the shorter-range oscillations as in the undoped case [Figs.~\ref{fig2}(c)-(d)] and a prevailing period $\pi/k_{F}$ that increases with $\mathcal{A}$.

In the following we derive the approximate analytical results for the RKKY coupling up to $\mathcal{O}(\Delta^2)$ for different impurity configurations when $2E_{F}>\Delta$.
The total exchange coupling can be written as a sum of the intrinsic contribution already obtained for undoped graphene and a second contribution due to a finite Fermi energy, $J({\bm R})=J_{{\rm int}}({\bm R})+J_{{\rm ext}}({\bm R})$. We start with the $AA$ zigzag configuration. The intrinsic contribution $J_{{\rm int}}({\bm R})$ is given by Eq.~\eqref{exa1}. To evaluate the extrinsic contribution, for convenience we define $y=\alpha_{\Delta}\sqrt{|(2\omega/\Delta)^2- 1|}$ and write $J_{{\rm ext}}({\bm R})=\hat{J}_{1}({\bm R}) +\bar{J}_{1}(\bm R)$,
where $\hat{J}_{1}({\bm R})$ and $\bar{J}_{1}(\bm R)$ take the following form
\begin{eqnarray}\label{ane1}
&&\hat{J}_{1}({\bm R})=\frac{4}{\pi}J_{+}F_0^2 \left(\frac{a}{R}\right)^3\int_{0}^{\alpha_{\Delta}}\BK(y)\BI(y)y\sqrt{y^2+\alpha_{\Delta}^2}dy\nonumber\\
&&+\frac{4}{\pi}J_{-} \left(\frac{ a}{2 l_{\Delta} }\right)^2\left(\frac{a}{R}\right)\int_{0}^{\alpha_{\Delta}}\BK(y)\BI(y)y(y^2+\alpha_{\Delta}^2)^{-\frac{1}{2}}dy\;,\nonumber\\
\end{eqnarray}
and
\begin{eqnarray}
&&\bar{J}_{1}({\bm R})=2J_{+}F_0^2 \left(\frac{a}{R}\right)^3\int^{k_F R}_{\alpha_{\Delta}}\BY(y)\BJ(y)y\sqrt{y^2+\alpha_{\Delta}^2}dy \nonumber \\
  &&+2J_{-} \left(\frac{ a}{2 l_{\Delta} }\right)^2\left(\frac{a}{R}\right) \int^{k_F R}_{\alpha_{\Delta}}y(y^2+\alpha_{\Delta}^2)^{-\frac{1}{2}}\BY(y)\BJ(y)dy,\nonumber\\\label{okf}
\end{eqnarray}
where the light-induced length scale $\l_{\Delta}$ is given by Eq.~\eqref{ani1}. Similar arguments for Eqs.~\eqref{ani1}-\eqref{Jpm1} applies here. Expansions of the integrands in Eq.~\eqref{ane1} up to second order in $y <\alpha_{\Delta}\ll 1$ show that both integrals are $\sim (\Delta/2)^3$ and are negligible. Eq.~\eqref{okf} for $\bar{J}(\bm R)$ can be evaluated by extending the  lower integration limit to $0$ with an negligible error $\sim (\Delta/2)^3$. Then $J_{{\rm ext}}(\bm R) \approx \bar{J}_{1}(\bm R)$ can be written as
\begin{eqnarray}
J_{{\rm ext}}(\boldsymbol{R})=&&2J_+F_0^2 \left(\frac{a}{R}\right)^3\int^{k_F R}_{0}\BY(y)\BJ(y)(y^2+\alpha_{\Delta}^2/2)dy\nonumber\\
&&+2J_{-}\left(\frac{ a}{2 l_{\Delta} }\right)^2\left(\frac{a}{R}\right)\int^{k_F R}_{0}\BY(y)\BJ(y)dy.\nonumber\\
\end{eqnarray}
The above integrals can be performed analytically, yielding $J({\bm R})=J_{{\rm int}}({\bm R})+J_{{\rm ext}}({\bm R})$ as
\begin{eqnarray} \label{AAzigzagMeijer}
&&\frac{J(\boldsymbol{R})}{J_0}=-\frac{F_0^2}{8} \left(\frac{a}{R}\right)^3\left[1+\cos{(2{\bm K}\cdot \boldsymbol{R})}\right]\left[1+16 {\mathcal{M}}_1(k_FR)\right]\nonumber\\
&&+\frac{1}{2} \left(\frac{ a}{2 l_{\Delta} }\right)^2\left(\frac{a}{R}\right)\left[3-\cos{(2{\bm K}\cdot \boldsymbol{R})}\right]\left[1-2 {\mathcal{M}}_2(k_FR)\right],\nonumber\\
\end{eqnarray}
where
\begin{eqnarray}
{\mathcal{M}}_1(x)&=&\frac{x}{2\sqrt{\pi}}{G}_{2,4}^{2,1} \left(\begin{matrix}&\frac{1}{2},&\frac{3}{2}& \\
                                       1,&1,&1,&-\frac{1}{2}
                          \end{matrix} \bigg|x^2 \right),\\
{\mathcal{M}}_2(x)&=&\frac{x}{2\sqrt{\pi}}{G}_{2,4}^{2,1} \left(\begin{matrix}&\frac{1}{2},&\frac{1}{2}& \\
                                       0,&0,&0,&-\frac{1}{2}
                          \end{matrix} \bigg|x^2 \right),
\end{eqnarray}
and ${G}_{p,q}^{m,n}$ is the Meijer G-function \cite{gradshteyn2007}. As in the undoped case, we find that Eq.~\eqref{AAzigzagMeijer} agrees well with our numerical results up to  $R \lesssim F_{0} l_{\Delta}$ for $\mathcal{A} \leq 0.04$. A more familiar form of the RKKY interaction can be obtained by considering the asymptotic regime $k_{F}R\gg1$ when the impurity separation is long compared to the Fermi wavelength. From the asymptotic forms of the Meijer G-functions we find
\begin{eqnarray}
&&\lim_{x\rightarrow \infty} {\mathcal{M}}_1(x)=-\frac{1}{16}+\frac{\cos{(2x)}}{8\pi}+\frac{x \sin{(2x)}}{2\pi},\nonumber\\
&&\lim_{x\rightarrow \infty}{\mathcal{M}}_2(x)=\frac{1}{2}-\frac{3\cos{(2x)}}{8\pi x^2}+\frac{\sin{(2x)}}{2\pi x},
\end{eqnarray}
from which we obtain
\begin{eqnarray}\label{asym1}
\frac{J(\boldsymbol{R})}{J_0}=&&-\frac{F_0^2}{\pi}\left(\frac{a}{R}\right)^3\left[1+\cos{(2\boldsymbol{K}\cdot \boldsymbol{R})}\right]\nonumber\\
&&\times\left[\frac{1}{4}\cos{(2k_F R)}+(k_F R)\sin{(2k_F R)}\right]\nonumber \\
&&+\frac{1}{2\pi} \left(\frac{ a}{2 l_{\Delta} }\right)^2\left(\frac{a}{R}\right)[3-\cos{(2\boldsymbol{K}\cdot\boldsymbol{R})}]\nonumber\\
&&\times\left[\frac{3\cos{(2k_F R)}}{4(k_F R)^2}-\frac{\sin{(2k_F R)}}{(k_F R)}\right].
\end{eqnarray}
\begin{figure*}
    \begin{center}
            \includegraphics[width=\textwidth]{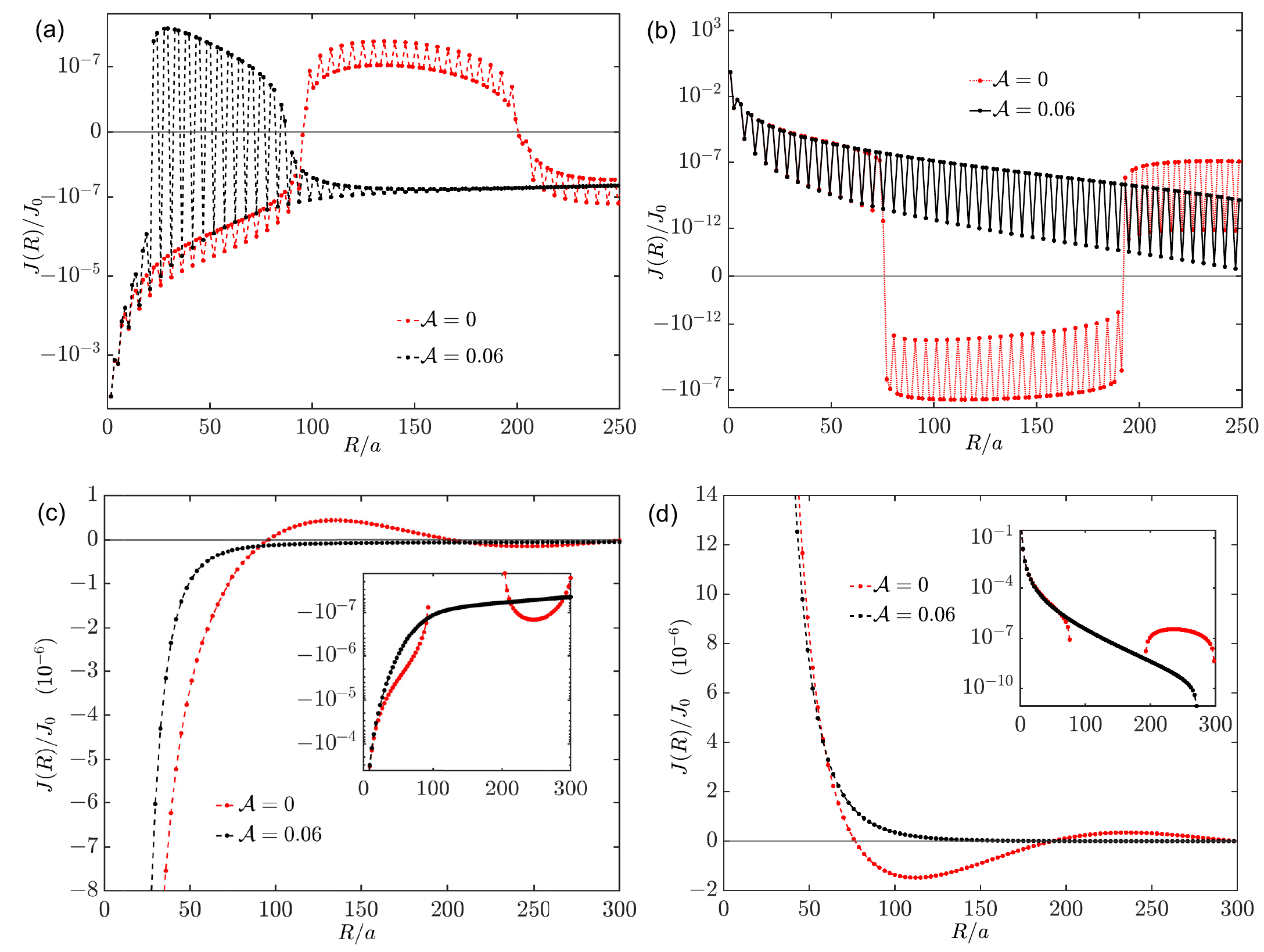}
                \end{center}
                \caption{Time-averaged RKKY coupling $J(R) /J_{0}$ for doped graphene under irradiation with a Fermi energy $E_{F}=0.0236\gamma$. The driving field has the same frequency and strengths as in Fig.~\ref{fig3}. For $\mathcal{A} = 0.06$ the photon-induced gap $\Delta=1.99 E_{F}$. The panels correspond to cases with the impurities located at different sublattice sites ($A$ or $B$) and separated along different directions (zigzag or armchair): (a) $AA$ zigzag; (b) $AB$ zigzag; (c) $AA$ armchair; (d) $AB$ armchair.
The insets in (c) and (d) show the negative and positive values of $J(R)/J_{0}$ respectively over the full range of $R$ down to the smallest separation. }
\label{fig4}
\end{figure*}
At equilibrium, the dominant contribution in $J(\boldsymbol{R})$ for finite Fermi energy goes as $1/R^2$ similar to 2D metals~\cite{satpathy1,serami,blackshaffer}, in contrast to the $1/R^3$ dependence in undoped graphene. Under irradiation, the doped case has a notable distinction from the undoped case, the dominant $R$-dependence in the second purely non-equilibrium term of  Eq.~\eqref{asym1} decays more rapidly going as $1/R^2$, as compared to $1/R$ in Eq.~\eqref{exa1}. As a result, the effect of this term in suppressing the exchange coupling becomes less important in the doped case. From Eq.~\eqref{asym1} we can clearly distinguish the lattice, Fermi surface, and irradiation effects at play in the system. The lattice effect is reflected by the period of short-range oscillations $\pi/\vert {\bm K} \vert$ and is not influenced by irradiation. On the other hand, the Fermi surface effect is manifested via the longer period $\pi/k_{F}$,
which is dependent on $\Delta$ and increases with the driving strength $\mathcal{A}$.

If we keep increasing the driving strength for a fixed $R$ we can leave the asymptotic regime and enter into a regime with $k_{F}R$ becomes smaller than $1$, since $k_{F}$ decreases with $\mathcal{A}$. This can be attained while keeping $\mathcal{A} \ll 1$.
Thus we can consider the limit $k_{F}R\ll1 $ and use the small argument expansions of the Meijer G-functions to obtain,
\begin{eqnarray}
&&\lim_{x\rightarrow 0} {\mathcal{M}}_1(x)=0,\nonumber\\
&&\lim_{x\rightarrow 0}{\mathcal{M}}_2(x)=-\frac{2x}{\pi}[1+\gamma+\ln{(x/2)}],
\end{eqnarray}
where $\gamma$ is the Euler-Mascheroni constant. Using the above, we find that Eq.~\eqref{AAzigzagMeijer} reduces to the same form as Eq.~\eqref{exa1} for irradiated undoped graphene.
Physically,  when the electrochemical potential is maintained constant in the system, increasing the driving strength $\mathcal{A}$ allows to ``drain out'' the Fermi sea via the increase in gap size $\Delta$.
To shed light further on this point, we draw a comparison between the two cases with the Fermi level far above the band gap [Fig.~\ref{fig3}(a)] and just above the band gap [Fig.~\ref{fig4}(a)].
Fig.~\ref{fig3}(a) shows, for a fixed $E_F$, the non-equilibrium and the equilibrium cases. Their difference is relatively small since the Fermi momentum $k_F$ under irradiation is still close to the equilibrium value when $E_F$ is large compared to $\Delta/2$. The difference between the irradiated and equilibrium cases is most dramatically illustrated when $E_F$ is fixed at a value just above the photon-induced gap under irradiation. Fig.~\ref{fig4}(a) shows $J(\bm{R})$ for such a value of $E_F$ in and out of equilibrium. The significant difference is due to the qualitatively distinct scenarios realized by increasing $\mathcal{A}$. When undriven, the RKKY behavior is that of doped graphene in equilibrium, which displays long-range RKKY oscillations with a period $\pi/k_{F}$. Under a driving field that induces a value of
$\Delta \lesssim 2E_F$, the Fermi surface is drastically reduced and the RKKY coupling approaches that of irradiated undoped graphene [Fig.~\ref{fig2}(a)].
This drastic change in the behavior of the RKKY exchange interaction under irradiation reveals a strong qualitative difference between the equilibrium and the irradiated cases; furthermore, it implies that by tuning the laser amplitude and frequency one can switch between the two qualitatively distinct regimes of RKKY interactions for doped and undoped graphene under irradiation.

One can follow similar calculation steps as outlined above for the $AA$ zigzag case to derive the analytical results for the $AB$ zigzag, and the $AA$ and $AB$ armchair configurations. In the asymptotic regime $k_{F}R\gg 1$, we find, for the $AB$ zigzag case,
\begin{eqnarray}\label{asy2}
&&\frac{J(\boldsymbol{R})}{J_0}=[1-\cos{(2{\bm K}\cdot {\bm R})}]\Bigg\{
\\&&\frac{F_0^2}{\pi}\left(\frac{a}{R}\right)^3 \left[\frac{5}{4}\cos{(2k_F R)}+k_F R\sin{(2k_F R)}\right]\nonumber\\
&&\left.-\left(\frac{ a}{2 l_{\Delta} }\right)^2 \left(\frac{a}{R}\right) \left[\frac{\cos{(2k_F R)}+4k_F R\sin{(2k_F R)}}{8\pi (k_F R)^2}\right]\right\},\nonumber
\end{eqnarray}
whereas for the $AA$ armchair case,
\begin{eqnarray}\label{asy3}
  &&\frac{J(\boldsymbol{R})}{J_0}=-\frac{F_0^2}{2\pi} \left(\frac{a}{R}\right)^3 \left[\cos{(2k_F R)}+4k_F R\sin{(2k_F R)}\right]\nonumber\\
&&+\frac{1}{4\pi} \left(\frac{ a}{2 l_{\Delta} }\right)^2\left(\frac{a}{R}\right) \left[\frac{3\cos{(2k_F R)}}{(k_F R)^2}-\frac{4\sin{(2k_F R)}}{k_F R}\right],\nonumber\\
\end{eqnarray}
and for the $AB$ armchair case,
\begin{eqnarray}
  &&\frac{J(\boldsymbol{R})}{J_0}=\frac{F_0^2}{\pi} \left(\frac{a}{R}\right)^3 \left[\frac{5}{2}\cos{(2k_F R)}+2k_F R\sin{(2k_F R)}\right]\nonumber\\
&&- \left(\frac{ a}{2 l_{\Delta} }\right)^2\left(\frac{a}{R}\right)\left[\frac{\cos{(2k_F R)}+4k_F R\sin{(2k_F R)}}{4\pi (k_F R)^2}\right].\nonumber\\
\end{eqnarray}
For all these cases, we also find from Eq.~\eqref{AAzigzagMeijer} that the $k_{F}R\ll 1$ behavior of $J(\boldsymbol{R})$ reduces to their corresponding results in the undoped regime under irradiation  [$AB$-zigzag, Eq.~\eqref{exa2}; $AA$-armchair, Eq.~\eqref{exa3}; $AB$-armchair, Eq.~\eqref{exa4}]. The underlying reason is already explained above as in the $AA$ zigzag case.
\begin{figure}
            \includegraphics[width=1\columnwidth]{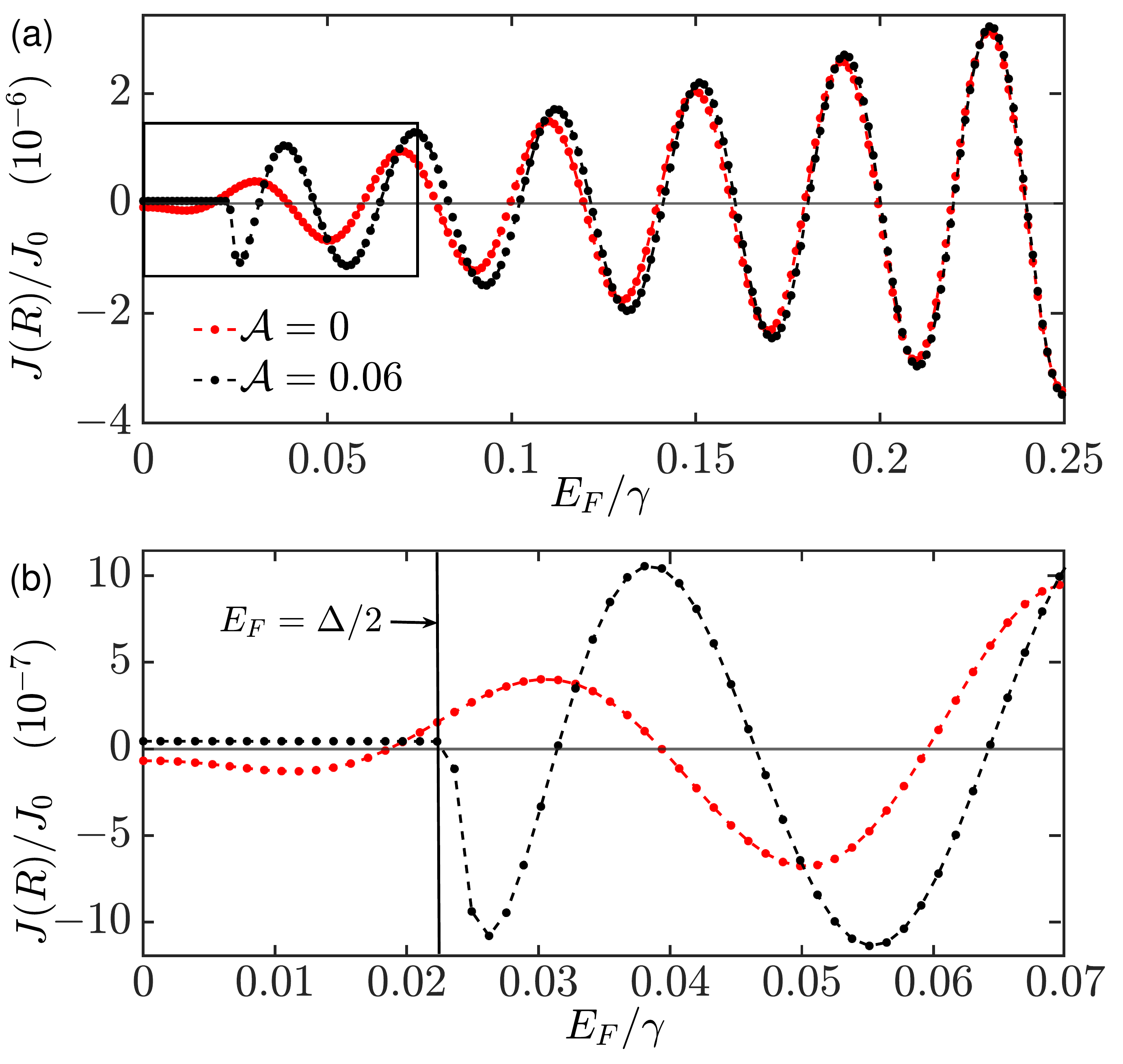}
            \caption{(a) Dependence of $J(\bm R)/J_0$ on the Fermi energy. In this case we consider the impurity spins to be separated by a fixed distance $R=70\sqrt{3}a$ on the same sublattice sites  $AA$ along the zigzag direction.  Panel (b) shows an enlarged view of  the small Fermi energy region marked by the rectangle in (a). The position of the conduction band edge at $E_F = \Delta/2$ is indicated by the vertical line. The driving frequency is the same as in Figs.~\ref{fig2}-\ref{fig3}.}
\label{fig5}
\end{figure}

The effect of the Fermi energy on the non-equilibrium RKKY interaction can be further elucidated by plotting $J(\bm{R})$ as a function of $E_F$ at a fixed $R$, as shown in Fig.~\ref{fig5}(a). It becomes clear that the RKKY coupling under irradiation approaches the equilibrium case when the Fermi energy is far from the conduction band edge $E_{F}\gg \Delta/2$. Departure from the equilibrium case occurs most considerably in the region of low Fermi energies, which is enlarged as shown in Fig.~\ref{fig5}(b). In particular, one notices that $J(\bm{R})$ in the irradiated case remains  constant throughout $E_F \in [0, \Delta/2]$, because the Fermi level stays within the photon-induced gap. Thus at a fixed $E_F > 0$, as one increases the driving strength $\mathcal{A}$ from zero, the gap $\Delta$ increases, and the RKKY coupling changes from having a predominantly intravalley scattering character to having a purely intervalley scattering character. For large enough $\mathcal{A}$ when the dynamical gap encloses the Fermi level,
the RKKY coupling remains constant behaving as irradiated undoped graphene.

\section{Discussion and outlook}\label{Discussion}

Our theory applies for dilute concentration of magnetic impurities located on the graphene sublattice sites. One interesting direction for future consideration is the inclusion of spin-orbit coupling effects.
For sufficient concentration of magnetic adatoms (\textit{e.g.} transition metal elements), the band structure of graphene becomes strongly modified due to the induced spin-orbit coupling and Zeeman splitting \cite{grapQAHE1,grapQAHE2,ASm1,Asm2}.  Our formalism can be further generalized to account for the induced spin-orbit coupling in the graphene electrons.

In equilibrium, it has been known that the RKKY interaction in undoped graphene is antiferromagnetic for impurity spins located at the same sublattice sites and ferromagnetic at different lattice sites. This statement is general for bipartite lattices
and is satisfied due to the particle-hole symmetry in undoped graphene \cite{serami}. In the presence of a driving field, one surprising finding from our study is the sign change of $J(\bm R)$ observed for the $AA$ zigzag case when $\mathcal{A} = 0.06$ [Fig.~\ref{fig2}(a)], resulting in a ferromagnetic coupling over a wide range of $R$. This is also depicted in Fig.~\ref{fig4}(a), in which the $\mathcal{A} = 0.06$ case exhibits essentially the same behavior as undoped graphene due to the small Fermi surface resulting from a large induced dynamical gap. Within our analytic approach, we have shown that this is due to competition between the equilibrium-like antiferromagnetic contribution and a purely light-induced ferromagnetic contribution [Eq.~\eqref{exa1}]. When $\mathcal{A}$ is large enough, the latter term wins out. An open question for future investigation would be to fully examine this sign change under larger values of driving strengths beyond the weak drive regime.

In doped system, we have demonstrated that the oscillation period of the RKKY coupling can be tuned by the driving field.  This tunability is due to the dynamical band gap opened at the Dirac point under circularly polarized light and is more enhanced in gapless systems like graphene. One can compare this to the RKKY interaction in intrinsically gapped system such as doped III-V semiconductors. Because of its large intrinsic gap, the band gap increase due to renormalization by optical Stark effect under a subgap frequency irradiation will be relatively small, and the tunability of its RKKY interaction will be less dramatic. The qualitatively distinct behaviors of the RKKY coupling for doped and undoped irradiated graphene, marked by the presence and absence of long-range oscillations respectively, pose a tantalizing prospect for optical switching of ferromagnetic/antiferromagnetic coupling. Starting from slightly doped graphene at equilibrium, turning on the laser can switch the coupling from antiferromagnetic to ferromagnetic for the $AA$ zigzag case [Fig.~\ref{fig4}(a)] in the impurity separation range of $R/a \sim 100\,-\,200$, and from ferromagnetic to antiferromagnetic for the $AB$ zigzag case [Fig.~\ref{fig4}(b)] in $R/a \sim 70\,-\,200$. Similar sign change can also be seen for the $AA$ and $AB$ armchair cases [Fig.~\ref{fig4}(c)-(d)]. This observation suggests further investigation into Floquet control of the magnetic ordering in magnetically doped graphene would be promising.

\section{Conclusion}\label{conclusion}

This work has presented a formalism for calculating the RKKY interaction in graphene irradiated by a circularly polarized light.  We have obtained a generic expression of the time-averaged RKKY coupling in terms of Keldysh-Floquet Green's functions. By transforming the original Floquet Hamiltonian into a new basis, we arrived at a new Hamiltonian that carries a transparent physical meaning as a tight-binding-like Hamiltonian in the Floquet space, allowing for a systematic order-by-order inclusion of different photon processes. In this work we focus on the off-resonant, weak drive regime where  only  virtual photon absorptions and emissions happen, which allow us to truncate terms corresponding to first and higher order photon processes. In this framework we have obtained the numerical and approximate analytical results for the time-averaged RKKY coupling under a combinations of different impurity locations and separation directions. When the Fermi level is at the Dirac point (undoped graphene), we find that the magnitude of the RKKY coupling is decreased by increasing irradiation strength while maintaining its ferromagnetic or antiferromagnetic character. When the driving field becomes strong enough, our results suggest that the ferromagnetic coupling in the case of impurities at the same sublattice sites and separated along the zigzag direction can be switched to antiferromagnetic coupling for a wide range of impurity separation. When the Fermi level is above the Dirac point (doped graphene), the RKKY coupling is characterized by long-range oscillations with a period that can be tuned by the driving field. For large values of driving strength, the RKKY behavior of doped graphene turns into that of undoped graphene through the optically induced metal-insulator transition.

\acknowledgments {This work was supported by the U.S. Department of Energy, Office of Science, Basic Energy Sciences under Early Career Award No. DE-SC0019326.}

\appendix
\section{Derivation of the exchange energy}\label{apexchf0}
In this appendix we provide a detailed derivation of the formulas for the non-equilibrium exchange energy in Eqs.~\eqref{exch1}-\eqref{exchJ}. We start by substituting Eq.~\eqref{Dyson} into Eq.~\eqref{exch} and obtain,
\begin{eqnarray}\label{exchtime}
&&E_{\alpha\beta}(\boldsymbol{R},t)=- i \lambda^2 \sum_{\mu,a,b,\nu} (S_{\mu})_{a b}  (S_{\nu})_{b a} \mathbb{S}_{1,\mu} \mathbb{S}_{2,\nu}
\int_{-\infty}^{+\infty} d t'\nonumber \\
&&  [G_{\alpha\beta} ^{\rm R}(\boldsymbol{R},t,t')G_{\beta\alpha} ^{<}(-\boldsymbol{R},t',t)+G^{<} _{\alpha\beta}(\boldsymbol{R},t,t')G_{\beta\alpha} ^{\rm A}(-\boldsymbol{R},t',t)].\nonumber\\
\end{eqnarray}
The time-dependent Green's functions above can first be transformed into the Floquet representation and then expressed as a Fourier series \cite{Oka_RMP},
\begin{eqnarray}
G^{\rm R}(\boldsymbol{R},t,t') =&&\sum_{m,n}\int^{\frac{\Omega}{2}}_{-\frac{\Omega}{2}}{\frac{d\bar{\omega}}{2\pi}}\\ &&\times e^{-i(\bar{\omega}+m\Omega) t+i(\bar{\omega}+n\Omega) t'}[G^{\rm R}(\boldsymbol{R},\bar{\omega})]_{mn},\nonumber
\end{eqnarray}
Substituting this Floquet representation of the Green's function, Eq. \eqref{exchtime} becomes
\begin{eqnarray}
&&E_{\alpha\beta}(\boldsymbol{R},t)=\nonumber\\
&&- i \lambda^2 \sum_{\mu,a,b,\nu} (S_{\mu})_{a b}  (S_{\nu})_{b a} \mathbb{S}_{1,\mu} \mathbb{S}_{2,\nu}
\int_{-\infty}^{+\infty} d t'\int_{-\frac{\Omega}{2}}^{\frac{\Omega}{2}}\frac{d\bar{\omega} d\bar{\omega}'}{(2\pi)^2}\nonumber \\
&&\sum_{m,n,m',n'}e^{-i[(\bar{\omega}-\bar{\omega}')+(m-m')\Omega]t}
e^{i[(\bar{\omega}-\bar{\omega}')+(n-n')\Omega]t'}\nonumber\\
&& \left\{ [G_{\alpha\beta} ^{\rm R}(\boldsymbol{R},\bar{\omega})]_{mn}[G_{\beta\alpha} ^{<}(-\boldsymbol{R},\bar{\omega}')]_{n'm'}\right.\nonumber\\
&&\left.+[G^{<} _{\alpha\beta}(\boldsymbol{R},\bar{\omega})]_{mn}[G_{\beta\alpha} ^{\rm A}(-\boldsymbol{R},\bar{\omega}')]_{n'm'}\right\}.
\end{eqnarray}
Then we can carry out the integral over $t'$ to obtain
\begin{eqnarray}
&&E_{\alpha\beta}(\boldsymbol{R},t)=- i \lambda^2 \sum_{\mu,a,b,\nu} (S_{\mu})_{a b}  (S_{\nu})_{b a} \mathbb{S}_{1,\mu} \mathbb{S}_{2,\nu}\int_{-\frac{\Omega}{2}}^{\frac{\Omega}{2}}\frac{d\bar{\omega} d\bar{\omega}'}{(2\pi)^2}\nonumber \\
&&\sum_{m,n,m',n'}e^{-i[(\bar{\omega}-\bar{\omega}')+(m-m')\Omega]t}2\pi\delta[(\bar{\omega}-\bar{\omega}')+(n-n')\Omega]\nonumber\\
&& \left\{ [G_{\alpha\beta} ^{\rm R}(\boldsymbol{R},\bar{\omega})]_{mn}[G_{\beta\alpha} ^{<}(-\boldsymbol{R},\bar{\omega}')]_{n'm'}\right.\nonumber\\
&&\left.+[G^{<} _{\alpha\beta}(\boldsymbol{R},\bar{\omega})]_{mn}[G_{\beta\alpha} ^{\rm A}(-\boldsymbol{R},\bar{\omega}')]_{n'm'}\right\}.
\end{eqnarray}
The result of $\bar{\omega}'$ integral above is nonzero only when $n=n'$ as both $\bar{\omega}$ and $\bar{\omega}'$ has to be in the reduced zone. Therefore,
\begin{eqnarray}
&&E_{\alpha\beta}(\boldsymbol{R},t)=- i \lambda^2 \sum_{\mu,a,b,\nu} (S_{\mu})_{a b}  (S_{\nu})_{b a} \mathbb{S}_{1,\mu} \mathbb{S}_{2,\nu}\nonumber \\
&&\int_{-\frac{\Omega}{2}}^{\frac{\Omega}{2}}\frac{d\bar{\omega} }{2\pi}\sum_{m,n,m'}e^{-i[(\bar{\omega}-\bar{\omega}')+(m-m')\Omega]t}\nonumber\\
&&\left\{ [G_{\alpha\beta} ^{\rm R}(\boldsymbol{R},\bar{\omega})]_{mn}[G_{\beta\alpha} ^{<}(-\boldsymbol{R},\bar{\omega})]_{nm'}\right.\nonumber\\
&&\left.+[G^{<} _{\alpha\beta}(\boldsymbol{R},\bar{\omega})]_{mn}[G_{\beta\alpha} ^{\rm A}(-\boldsymbol{R},\bar{\omega})]_{nm'}\right\},
\end{eqnarray}
and finally we take the time average,
\begin{eqnarray}
&&E_{\alpha\beta}(\boldsymbol{R})=\frac{1}{T}\int_0^{T}dt E_{\alpha\beta}(\boldsymbol{R},t)\\
&&=- i \lambda^2 \sum_{\mu,a,b,\nu} (S_{\mu})_{a b}  (S_{\nu})_{b a} \mathbb{S}_{1,\mu} \mathbb{S}_{2,\nu}\int_{-\frac{\Omega}{2}}^{\frac{\Omega}{2}}\frac{d\bar{\omega} }{2\pi}\sum_{m,n,m'}\delta_{m,m'}\nonumber\\
&&\left\{ [G_{\alpha\beta} ^{\rm R}(\boldsymbol{R},\bar{\omega})]_{mn}[G_{\beta\alpha} ^{<}(-\boldsymbol{R},\bar{\omega})]_{nm'}\right.\nonumber\\
&&\left.+[G^{<} _{\alpha\beta}(\boldsymbol{R},\bar{\omega})]_{mn}[G_{\beta\alpha} ^{\rm A}(-\boldsymbol{R},\bar{\omega})]_{nm'}\right\},
\end{eqnarray}
and then sum over $m'$ to arrive at the exchange in the Floquet representation in Eq.~\eqref{exch1}.

From the definitions of the real-space Green's functions \cite{Jauho_book}, one can show that $[G^{\rm A}(-{\bm R})]^{\dag} =G^{\rm R}(\boldsymbol{R})$ and $[G^<(\boldsymbol{R})]^{\dagger} = -[G^<(-\boldsymbol{R})]$, which implies $G_{\beta\alpha}^{\rm A}(-\boldsymbol{R})^{*} = G_{\alpha \beta}^{\rm R}(\boldsymbol{R})$ and $G_{\beta \alpha}^<({\bm R})^{*} = -G_{ \alpha\beta}^<(-{\bm R})$.
These relations can also be seen explicitly generically from the momentum-space Green's functions, \textit{e.g.},
\begin{eqnarray}
[G^<(\boldsymbol{R})]^{\dagger}&=&\left[\int d\bk e^{i \bk \cdot {\bm R}} G^{\rm R}_{\bk}\Sigma^< G^{\rm A}_{\bk} \right]^{\dagger}\nonumber
\\
&=&\int d\bk e^{-i \bk \cdot{\bm R}}  G^{\rm R}_{\bk}(\Sigma^<)^{\dagger} G^{\rm A}_{\bk}\nonumber
\\
&=&-\int d\bk e^{-i \bk \cdot{\bm R}} G^{\rm R}_{\bk}\Sigma^< G^{\rm A}_{\bk}\nonumber
\\
&=&-[G^<(-\boldsymbol{R})]\;.
\end{eqnarray}
With these relations we can write the integrand in Eq.~\eqref{exch1} as
\begin{eqnarray} \label{ImGG1}
&\textrm{Tr}\{G^{\rm R}_{\alpha \beta}(\boldsymbol{R},\bar{\omega})G^<_{ \beta\alpha}(-\boldsymbol{R},\bar{\omega})+G^<_{\alpha \beta}(\boldsymbol{R},\bar{\omega})G^{\rm A}_{\beta\alpha }(-\boldsymbol{R},\bar{\omega})\}\nonumber
\\
&=2i \Im{\textrm{Tr}\{G^{\rm R}_{\alpha \beta}(\boldsymbol{R},\bar{\omega})G^<_{ \beta\alpha}(-\boldsymbol{R},\bar{\omega})\}},
\end{eqnarray}
which leads to Eq.~\eqref{exchJ}.


\section{Unitary transformations for the photon-number representation}\label{unittrans}
In this appendix we describe the set of unitary transformations used to obtain the Hamiltonian in the photon number representation $\tilde{H}_{F}$ in Eq.~\eqref{us}~\cite{busl}.
The first transformation rotates the original Floquet Hamiltonian $H_{F}$ at $\bk=0$ so that it becomes block diagonal.
At the $K$ point this unitary transformation takes the form
\begin{eqnarray}
&U_{1,\tau=1}=
&\begin{bmatrix}
\ddots&&&&
\\ &\sigma_x&0&0&
\\&0&\sigma_x&0&
\\&0&0&\sigma_x&
\\&&&&\ddots
 \end{bmatrix},
\end{eqnarray}
while for the $K'$ point this transformation takes the form $U_{1,\tau=-1}=\mathbb{I}_{\infty}$, where $\mathbb{I}_{\infty}$ is identity operator in the Floquet space.

The second transformation is constructed as the transformation that diagnonalizes the transformed Floquet Hamiltonian $U_{1,\tau}^{\dag}H_{F}U_{1,\tau}$ at $k = 0$. If we denote the spinor eigenvectors corresponding to the eigenvalues $\epsilon_{n,\pm}=\pm\Delta/2+n\hbar\Omega$ as ${\phi}^{\tau}_{n,\pm}$, then
  $U_{2,\tau}=[\ldots, {\phi}^{\tau}_{n-1,+},{\phi}^{\tau}_{n-1,-},{\phi}^{\tau}_{n,+},{\phi}^{\tau}_{n,-},{\phi}^{\tau}_{n+1,+},{\phi}^{\tau}_{n+1,-},\ldots]$, where $\Delta$ is given in Eq.~\eqref{hf0} and $\tau=+$ ($\tau=-$) corresponds to the $K$ ($K'$) point. This transformation takes the explicit form
\begin{eqnarray}
&&U_{2,\tau=\pm1}=\nonumber
\\
&&\small\begin{bmatrix}
\ddots&&&&
\\ &g_+P_{A}\pm ih_-P_{B}&\pm ih_+\sigma_+&0&
\\&\sigma_-g_-&g_+P_{A}\pm ih_-P_{B}&\pm ih_+\sigma_+&
\\&0&\sigma_-g_-&g_+P_{A}\pm ih_-P_{B}&
\\&&&&\ddots
 \end{bmatrix},\nonumber\\
\end{eqnarray}
where we have defined the projection operator $P_{A, B}$ for projecting onto the $A$ or $B$ sublattice
\begin{eqnarray}\label{pab}
P_{A}=
\begin{bmatrix}
1&0\\
0&0
 \end{bmatrix},\,\,\,\,\,
P_{B}=
\begin{bmatrix}
0&0\\
0&1
 \end{bmatrix}.
\end{eqnarray}
Then we apply this transformation on the Hamiltonian $U_{1,\tau}^{\dag}H_{F}U_{1,\tau}$ for all $\bk$ to arrive at  $U^{\dag}_{2,\tau}U_{1,\tau}^{\dag}H_{F}U_{1,\tau}U_{2,\tau}$. Finally, in order to establish a scalar product structure between the isospin and momentum in the transformed Hamiltonian that mimics the equilibrium Hamiltonian of graphene, we apply the valley-independent unitary transformation $U_{3}$,
\begin{eqnarray}
&&U_{3}=\nonumber
\\
&&\begin{bmatrix}
\ddots&&&&
\\ &P_{A}-i P_{B}&0&0&
\\&0&P_{A}-i P_{B}&0&
\\&0&0&P_{A}-i P_{B}&
\\&&&&\ddots
 \end{bmatrix},
\end{eqnarray}
to arrive at the final transformed Hamiltonian $U^{\dag}_{3}U^{\dag}_{2,\tau}U_{1,\tau}^{\dag}H_{F}U_{1,\tau}U_{2,\tau}U_{3}=\tilde{H}_{F}$ in Eq.~\eqref{us}, which takes the explicit matrix form
\\
\\
\begin{widetext}
\begin{equation}\label{HFFtil}
 \tilde{H}_{F}= \left(
                  \begin{array}{cccccccc}
                    \ddots & \vdots & \vdots & \vdots & \vdots & \vdots & \vdots & \reflectbox{$\ddots$}\\
                    \ldots & \Delta/2-\hbar\Omega & e^{i\theta_{k}}F_{0}\hbar v_{F} k  &  -e^{i\theta_{k}}F_{1}\hbar v_{F} k & 0 & 0 & 0 & \ldots \\
                    \ldots &  e^{-i\theta_{k}}F_{0}\hbar v_{F} k & -\Delta/2-\hbar\Omega & 0 & e^{i\theta_{k}}F_{1}\hbar v_{F} k & e^{i\theta_{k}}F_{2}\hbar v_{F} k & 0 & \ldots \\
                    \ldots &  e^{-i\theta_{k}}F_{1}\hbar v_{F} k  & 0 & \Delta/2 &  e^{i\theta_{k}}F_{0}\hbar v_{F} k & -e^{i\theta_{k}}F_{1}\hbar v_{F} k & 0 & \ldots \\
                    \ldots & 0 & -e^{-i\theta_{k}}F_{1}\hbar v_{F} k &  e^{-i\theta_{k}}F_{0}\hbar v_{F} k & -\Delta/2 & 0 & e^{i\theta_{k}}F_{1}\hbar v_{F} k & \ldots \\
                    \ldots & 0 & e^{-i\theta_{k}}F_{2}\hbar v_{F} k  & e^{-i\theta_{k}}F_{1}\hbar v_{F} k & 0 & \Delta/2+\hbar\Omega &  e^{i\theta_{k}}F_{0}\hbar v_{F} k & \ldots \\
                    \ldots & 0 & 0 & 0 &  -e^{-i\theta_{k}}F_{1}\hbar v_{F} k &  e^{-i\theta_{k}}F_{0}\hbar v_{F} k & -\Delta/2+\hbar\Omega & \ldots \\
                   \reflectbox{$\ddots$} & \vdots & \vdots & \vdots & \vdots & \vdots & \vdots & \ddots \\
                  \end{array}
                \right)\;.
\end{equation}
\end{widetext}

\section{Floquet real-space Green's functions}\label{rsgf}
In this section we outline the steps used to derive the real-space Green's functions in Eqs.~\eqref{pm1}, \eqref{pm2}, \eqref{gl1} and \eqref{gl2}. In the Fourier transform of Eq.~\eqref{FourG}, we use the Jacobi–Anger formula to expand $e^{i \boldsymbol{k}\cdot\boldsymbol{R}}$:
\begin{eqnarray}
e^{i \boldsymbol{k}\cdot\boldsymbol{R}}=e^{i kR\cos{(\theta_R-\theta_k)} }=\sum_{n=-\infty}^{\infty}i^n \mathcal{J}_n(kR)e^{i n (\theta_R-\theta_k)}.\nonumber\\
\end{eqnarray}
 Integrating over the azimuthal angle for each of the components of the Green's function, one term of the Jacobi–Anger expansion survives due to orthogonality. For example considering $\uparrow\uparrow$ component of the retarded Green's function we get,
\begin{eqnarray}
&&[\tilde{g}^R(\boldsymbol{R},\omega)]_{\uparrow\uparrow}=\frac{1}{2\pi}\frac{1}{(\hbar v_F F_0)^2}\\
&&\times\int_{0}^{\infty} dk  k J_0(k R)\frac{\omega+i\eta+\Delta/2}{[(\omega+i\eta)^2-\Delta^2/4]/(\hbar v_F F_0)^2-k^2},\nonumber
\end{eqnarray}
The Bessel function $\BJ(k R)$ can be written as $\BJ(k R)=[{H}_0^{(1)}(k R)+{H}_0^{(2)}(k R)]/2=[{H}_0^{(1)}(k R)-{H}_0^{(1)}(-k R)]/2$, where ${H}^{(1)}$ (${H}^{(2)}$) is the Hankel function of the first (second) kind. Then by changing variables $k\rightarrow-k$ for the second term and simplifying we get the integral,
\begin{eqnarray}
I=\int_{-\infty}^{\infty} dk \frac{k {H}_0^{(1)}(k R)}{k^2-[(\omega+i\eta)^2-\Delta^2/4]/(\hbar v_F F_0)^2}.
\end{eqnarray}
For large $\vert z \vert$, the Hankel function ${H}_0^{(1)}$ behaves asymptotically as
\begin{eqnarray}
{H}_0^{(1)}(z)\approx\sqrt{\frac{2}{\pi}}\sqrt{\frac{1}{z}}e^{i(z-\frac{\pi}{4})},
\end{eqnarray}
for$-\pi<\arg{z}<2\pi$. Therefore, the magnitude of this function satisfies,
\begin{eqnarray}
\abs{{H}_0^{(1)}(z)}\approx\sqrt{\frac{2}{\pi}}\sqrt{\frac{1}{\abs{z}}}e^{-\abs{z}\sin(\arg{z})},
\end{eqnarray}
which goes to zero with $\abs{z} \to \infty$  in the upper half plane.

Thus the integral can be performed by completing the contour in the upper half of the complex plane and using the residue theorem. The contribution from the integral on the half circle vanishes as the magnitude of the Hankel function decays exponentially for $\abs{z} \to \infty$. To determine the poles in the complex plane we solve the equation $k^2-[(\omega+i\eta)^2-\Delta^2/4]/(\hbar v_F F_0)^2=0$, and use the pole in the upper half complex plane for the residue theorem.

We write out the principle value of the square root in explicit form,
\begin{eqnarray}
&&\sqrt{(\omega+i\eta)^2-\Delta^2/4}=
\\
&&\sqrt{\frac{\omega^2-\eta^2-\Delta^2/4+\sqrt{(\omega^2-\eta^2-\Delta^2/4)^2+4\eta^2\omega^2}}{2}}+
\nonumber\\&&\sqrt{\frac{\sqrt{(\omega^2-\eta^2-\Delta^2/4)^2+4\eta^2\omega^2}-(\omega^2-\eta^2-\Delta^2/4)}{2}}\nonumber\\
&&\times i\ {\rm sgn}(\omega).\nonumber
\end{eqnarray}
Among the two poles of the integrand $k = \pm \sqrt{(\omega+i\eta)^2-(\Delta/2)^2}/(\hbar v_F F_{0})$, we can observe that $\sqrt{(\omega+i\eta)^2-(\Delta/2)^2}/(\hbar v_F F_{0})$ is in the upper half plane with a positive imaginary part when $\omega>0$ and $-\sqrt{(\omega+i\eta)^2-(\Delta/2)^2}/(\hbar v_F F_{0})$ is in the upper half plane when $\omega<0$. Then with this procedure we then calculate the value of the residue at the pole and arrive at the result in the main text [Eq.~\eqref{pm1}] for the Green's function
\begin{equation}
[\tilde{g}^R(\boldsymbol{R},\tilde\omega)]_{\uparrow\uparrow }=-\zeta[\omega+i\eta+\Delta/2]\chi_0(R,\omega).
\end{equation}
The other components of the retarded Green's functions in real space are calculated in a similar way.

The momentum space representation of the lesser Green's function is,
\begin{widetext}
\begin{eqnarray}
\tilde{g}^{<}(\boldsymbol{k},\omega)=\frac{2i\eta F_0 f(\omega)}{D_+(k,\omega)D_-(k,\omega)}\begin{bmatrix}
   ( \omega+\Delta/2)^2+\eta^2+ (\hbar v_F F_0 k)^2 &  2\omega\hbar v_F F_0 k  e^{i \theta_k} \\
  2\omega\hbar v_F F_0 k  e^{-i \theta_k} &   ( \omega-\Delta/2)^2+\eta^2+(\hbar v_F F_0 k)^2
 \end{bmatrix}\;.
\end{eqnarray}
\end{widetext}
Then the for the $\uparrow\uparrow$ component we get
\begin{eqnarray}
&&[\tilde{g}^<(\boldsymbol{R},\omega)]_{\uparrow\uparrow}=\frac{2i\eta F_0 f(\omega)}{(2\pi)^2}\int d\boldsymbol{k} e^{i \boldsymbol{k}\cdot\boldsymbol{R}} [\tilde{g}^<(\boldsymbol{k},\omega)]_{\uparrow\uparrow}
\\
&&=\frac{2i\eta F_0 f(\omega)}{(2\pi)^2}\int d\boldsymbol{k} e^{i \boldsymbol{k}\cdot\boldsymbol{R}}\frac{ \left[ \omega+(\Delta/2)^2+\eta^2+(\hbar v_F F_0 k)^2\right]}{D_+(k,\omega)D_-(k,\omega)}\;.\nonumber
\end{eqnarray}
We decompose the integrand into partial fractions:
\begin{eqnarray}
&&\frac{k}{D_+(k,\omega)D_-(k,\omega)}=\frac{k}{4i\eta\omega}\left[\frac{1}{D_+(k,\omega)}-\frac{1}{D_-(k,\omega)}\right]\;,\nonumber\\
\end{eqnarray}
\begin{eqnarray}
&&\frac{k^3}{D_+(k,\omega)D_-(k,\omega)}\\
&&=\frac{1}{2}[1-B(\omega)]\frac{k}{D_+(k,\omega)}+\frac{1}{2}[1+B(\omega)]\frac{k}{D_-(k,\omega)}\;,\nonumber
\end{eqnarray}
so that each term can be integrated as in the previous calculation to get the result in the main text. The calculation of the other components of the lesser Green's functions follows a similar approach.

\section{RKKY coupling in the $F_0$ approximation}\label{RKKYF0}
In this appendix we outline the derivation of the non-equilibrium RKKY coupling Eq.~\eqref{exch2} within the $F_0$ approximation, and then describe how it can be reduced to the equilibrium result in the absence of a driving field.
We first transform the Green's functions in Eq.~\eqref{exchJ} from the original Floquet basis to the $K'$ photon-number basis within the $F_{0}$ approximation.
Let $\mathcal{P}_{\alpha}=P_{\alpha}\otimes \mathbb{I}_{\infty}$ be the projection operator onto the $\alpha$ sublattice in the combined pseudospin-Floquet Hilbert space, where $P_{\alpha}$ is given in Eq.~\eqref{pab}. It satisfies the following properties,
\begin{eqnarray}\label{pprop}
\mathcal{P}_{\alpha}\mathcal{P}_{\bar\alpha}=0,\,\,\,\,\,\mathcal{P}_{\alpha}\mathcal{P}_{\alpha}=\mathcal{P}_{\alpha},
\end{eqnarray}
where $\bar{\alpha}$ stands for the complement of $\alpha$, \textit{i.e.}, $\bar{\alpha}=B$ if $\alpha=A$, and vice versa.

Then the integrand in Eq.~\eqref{exchJ} can be written as
\begin{eqnarray}\label{new1}
I_J  &=&\Im{\textrm{Tr}\{G^R_{\alpha \beta}(\boldsymbol{R},\omega)G^<_{ \beta\alpha}(-\boldsymbol{R},\omega)\}}  \nonumber\\
  &=& \Im{\textrm{Tr}\{\mathcal{P}_{\alpha}G^{\rm R}(\boldsymbol{R},\omega)\mathcal{P}_{\beta}\mathcal{P}_{\beta}G^<(-\boldsymbol{R},\omega)\mathcal{P}_{\alpha}\}}\nonumber\\
     &=& \Im{\textrm{Tr}\{\mathcal{P}_{\alpha}G^{\rm R}(\boldsymbol{R},\omega)\mathcal{P}_{\beta}G^<(-\boldsymbol{R},\omega)\}},
\end{eqnarray}
where in the last line we made use of the cyclic property of the trace.
We can express the Green's functions in the Floquet representation in terms of the $K'$ photon-number basis:
\begin{eqnarray}
  G(\boldsymbol{R},\omega)=U_{T,-} \tilde{G}(\boldsymbol{R},\omega) U_{T,-}^{\dagger},
\end{eqnarray}
Then Eq.~\eqref{new1} can be expressed as
\begin{eqnarray}\label{new2}
  I_J &=&\Im\Big\{{\rm Tr}\{U^{\dag}_{T,-}\mathcal{P}_{\alpha}U_{T,-}\tilde{G}^{\rm R}(\boldsymbol{R},\omega)\nonumber\\
&& \times U^{\dag}_{T,-}\mathcal{P}_{\beta}  U_{T,-}\tilde{G}^<(-\boldsymbol{R},\omega)\}\Big\}.
\end{eqnarray}
Within the $F_{0}$ approximation we find that
\begin{eqnarray}\label{mapa}
&U_{T,-1}^{\dagger}\mathcal{P}_{\alpha}U_{T,-1}= F_0 \mathcal{P}_{\alpha}.
\end{eqnarray}
It is important to notice that even though the projection operator $\mathcal{P}_{\alpha}$ appears on both sides of Eq.~\eqref{mapa}, they have different physical meanings as they are expressed in different bases.
$\mathcal{P}_{\alpha}$ on the left hand side projects onto $\alpha$ component of the pseudospin basis of the original Floquet representation, while $\mathcal{P}_{\alpha}$ on the right projects onto the $\alpha$ component of the isospin basis of the $K'$ photon-number representation. Then substituting  Eq.~\eqref{mapa} in  Eq.~\eqref{new2} we get
\begin{equation}\label{new3}
  I_J = F^{2}_{0}\Im{{\rm Tr}\{\mathcal{P}_{\alpha}\tilde{G}^{\rm R}(\boldsymbol{R},\omega)\mathcal{P}_{\beta}\tilde{G}^<(-\boldsymbol{R},\omega)\}}.
\end{equation}
Using the projection operator property $\mathcal{P}_{\alpha}\mathcal{P}_{\alpha}=\mathcal{P}_{\alpha}$ in Eq.~\eqref{pprop} and the cyclic properties of the trace we can rewrite Eq.~\eqref{new3} as
\begin{equation}\label{new4}
I_J =F^{2}_{0}\Im{{\rm Tr}\{\tilde{G}^{\rm R}_{\alpha\beta}(\boldsymbol{R},\omega)\tilde{G}^<_{\beta\alpha}(-\boldsymbol{R},\omega)\}}.
\end{equation}
By replacing the integrand in Eq.~\eqref{exchJ} by Eq.~\eqref{new4} we obtain at Eq.~\eqref{exchJ2} in the main text. The Green's functions in Eq.~\eqref{exchJ2} are block-diagonal in the Floquet index $n$ and composed of the $2\times2$ Green's functions $\tilde{g}_{n}({\bm R, \bar{\omega}})$, hence the trace in Eq.~\eqref{exchJ2} results in a simple sum over $n$. By writing $\tilde{g}_{n}({\bm R, \bar{\omega}}) \equiv \tilde{g}({\bm R, \bar{\omega}-n\hbar\Omega})$ for each valley and using Eq.~\eqref{FourG3},
Eq.~\eqref{exchJ2} can be written as
\begin{eqnarray}\label{exchJ3}
  &&J_{\alpha \beta}({\bm R})= \lambda^2\hbar^2 F^2_{0} \sum_{n=-\infty}^{\infty}\int^{\frac{\hbar\Omega}{2}}_{-\frac{\hbar\Omega}{2}}\frac{d\bar{\omega} }{2\pi}\\ &&{\rm Im}\left\{[\tilde{g}_{F_0}^{\rm R}({\bm R},\bar{\omega}-n\hbar\Omega)]_{\alpha\beta}[\tilde{g}_{F_0}^<(-{\bm R},\bar{\omega}-n\hbar\Omega)]_{\beta\alpha}\right\}\;.\nonumber
\end{eqnarray}
Making the change of variable $\omega=\bar{\omega}-n\hbar\Omega$, the Green's functions in Eq.~\eqref{exchJ3} become independent of $n$ and the limits of integration become $[-(1/2+n)\hbar\Omega,(1/2-n)\hbar\Omega]$. Then, by performing the sum over the Floquet index $n$ we arrive at Eq.~\eqref{exch2} where $\omega\in (-\infty,\infty)$ in this equation is the extended zone frequency.

Next, we discuss how the exchange energy in Eq.~\eqref{exch2} reduces to its equilibrium counterpart in the absence of periodic driving.
In the equilibrium limit $\mathcal{A}\rightarrow 0$, $\Delta=0$, we have $F_{0}=1$ and $n=0$, so that the Hamiltonian in Eq.~\eqref{hf0} takes the form of $\tilde{\mathcal{H}}_{0}=\hbar v_{F}(\sigma_{x}k_{x}-\sigma_{y}k_{y})$ for both the $K$ and $K'$ points. Meanwhile, Eq.~\eqref{ghpm} gives $h_{-}=g_{+}=1$ and $h_{+}=g_{-}=0$, the isospin bases Eqs.~\eqref{bas1} and \eqref{bas2} become
$(\Phi_{\uparrow},\Phi_{\downarrow})^{T}=\sigma_{x}(\phi_{A},\phi_{B})^{T}$ for $K$ and $(\Phi_{\uparrow},\Phi_{\downarrow})^{T}=\sigma_{z}(\phi_{A},\phi_{B})^{T}$  for $K'$. Therefore, the equilibrium Hamiltonian $\tilde{\mathcal{H}}_{0}$  in the photon-number representation is related to that in the original pseudospin basis [Eq.~\eqref{h0tau}] by the following unitary transformations
\begin{equation}\label{unith}
\mathcal{H}_{0}=\sigma_{x}\tilde{\mathcal{H}}_{0}\sigma_{x}\; {\rm for\;} K\,\,\,\,\,{\rm and}\,\,\,\,\,\mathcal{H}_{0}=\sigma_{z}\tilde{\mathcal{H}}_{0}\sigma_{z}\;{\rm for\;} K'\;. 
\end{equation}
It follows that the equilibrium Green's functions in the photon-number representation are also related to those in the original pseudospin basis by
\begin{equation}\label{unitg}
g_{K} = \sigma_{x}\tilde{g}\sigma_{x}\; {\rm for\;} K\,\,\,\,\,{\rm and}\,\,\,\,\,g_{K'}=\sigma_{z}\tilde{g}\sigma_{z}\;{\rm for\;} K'\;. 
\end{equation}
In order to express Eq.~\eqref{exch2} in terms of the Green's functions $g_{K,K'}$ in the original pseudospin basis, we start by recalling that the Green's function in Eq.~\eqref{FourG3} is written in $K'$ isospin basis. Hence in equilibrium this Green's function transforms to the pseudospin basis as $\sigma_{z}\tilde{g}_{F_0}({\bm R},\omega)\sigma_{z}={g}({\bm R},\omega)$, such that
\begin{eqnarray}\label{FourG4}
  {g}({\bm R},\omega) &=& e^{i {\bm K}\cdot {\bm R}}\sigma_{z}\sigma_{y}\tilde{g}({\bm R},\omega)\sigma_{y}\sigma_{z}\nonumber\\
                                                                                                                    &&+e^{i {\bm K'}\cdot {\bm R}}\sigma_{z}\tilde{g}(\bm R,\omega)\sigma_{z}.
 \end{eqnarray}
 Since $\sigma_y\sigma_z = -\sigma_y\sigma_z = i\sigma_x$, the first term on the right in the above equation becomes $\sigma_{x}g_{F_0}\sigma_{x}$. Using Eq.~\eqref{unitg}, Eq.~\eqref{FourG4}  immediately reduces to its expected equilibrium form
\begin{eqnarray}\label{eqGF0}
{g}({\bm R},\omega)=e^{i {\bm K}\cdot {\bm R}}g_{K}({\bm R},\omega)+ e^{i {\bm K'}\cdot {\bm R}}g_{K'}({\bm R},\omega)\;.
\end{eqnarray}
Hence  Eq.~\eqref{exch2} in equilibrium becomes
\begin{eqnarray}\label{wellexc2}
J_{\alpha \beta}({\bm R})&=&\hbar^2\lambda^2\int^{\infty}_{-\infty}\frac{d\omega}{2\pi}\\&&{\rm Im}\left\{{g}_{\alpha\beta}^{{\rm R}}({\bm R},\omega) {g}_{\beta\alpha}^{{<}}(-{\bm R},\omega)\right\}\;.\nonumber
\end{eqnarray}
The above can be further simplified by making use of the fluctuation-dissipation theorem,
\begin{eqnarray}\label{fldiss}
{g}^{<}(-{\bm R},\omega)&=&-f(\omega)[{g}^{{\rm R}}(-{\bm R},\omega)-{g}^{\rm A}(-{\bm R},\omega)]\;.\nonumber\\
\end{eqnarray}
where $f(\omega)$ is the Fermi distribution function. By substituting Eq.~\eqref{fldiss} and using
${g}_{\beta\alpha}^{\rm A}(-{\bm R},\omega)={g}_{\alpha\beta}^{\rm R}({\bm R},\omega)^{*}$,
the integrand of Eq.~\eqref{wellexc2} becomes $-f(\omega){\rm Im}\{{g}_{\alpha\beta}^{{\rm R}}({\bm R},\omega) {g}_{\beta\alpha}^{R}(-{\bm R},\omega) \}$.
At low-temperatures where the Fermi distribution function $f(\omega)=\Theta(E_{F}-\omega)$, we arrive at the well-known form of the RKKY coupling~\cite{satpathy1}
\begin{equation}\label{wellexc}
J_{\alpha \beta}({\bm R})=-\hbar^2\lambda^2\int^{E_{F}}_{-\infty}\frac{d\omega}{2\pi}{\rm Im}\left\{{g}_{\alpha\beta}^{{\rm R}}({\bm R},\omega) {g}_{\beta\alpha}^{{\rm R}}(-{\bm R},\omega)\right\}.
\end{equation}

\section{Asymptotic behavior in the undoped case}\label{rsgf}

\begin{figure*}

   \begin{center}
            \includegraphics[width=\textwidth]{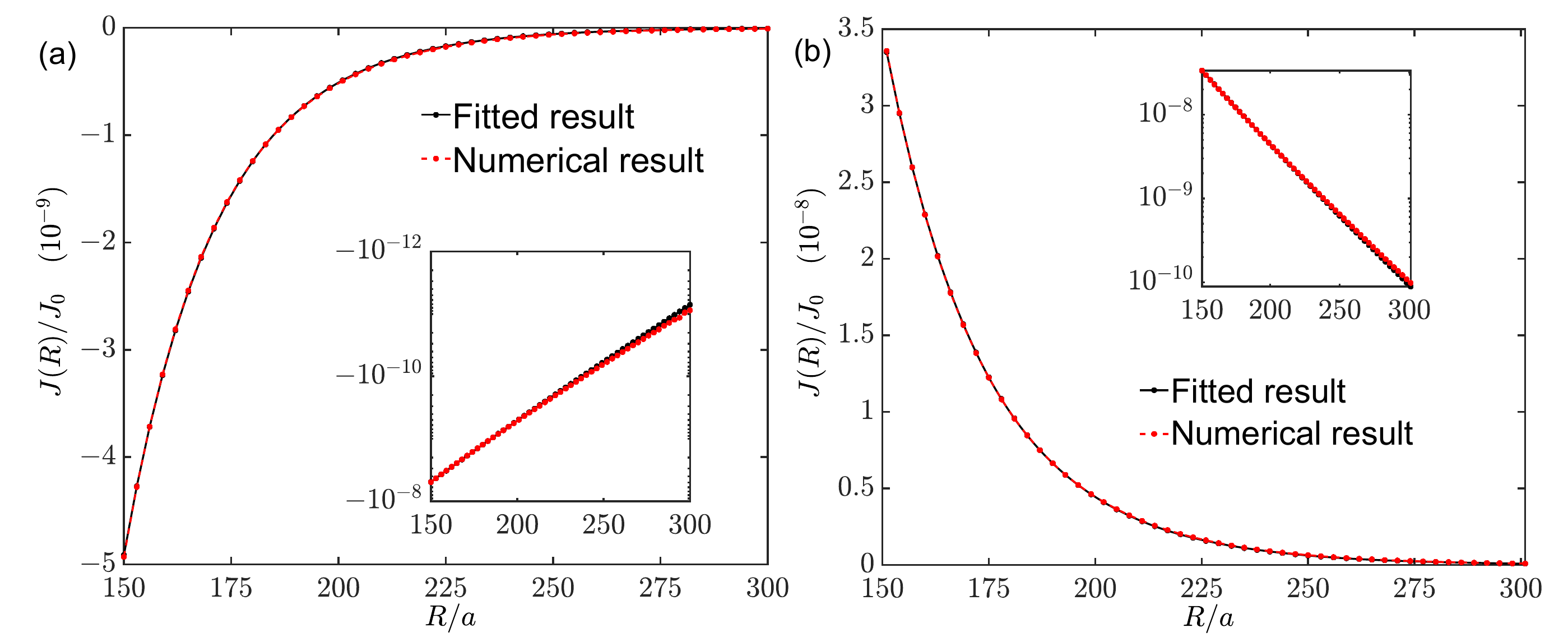}
                \end{center}
                \caption{Fitting result of $J(\bm R)/J_0$ with $\mathcal{A}=0.06$ and $E_F=0$ in the regime of $\alpha_{\Delta} \gtrsim 1$ for the cases of (a) AA armchair and (b) AB armchair.
$J_{0}$ is given in Eq.~\eqref{alpha}.}

\label{fig8}
\end{figure*}
We remark on the large-$R$ behavior of our non-equilibrium results for the undoped case, comparing with the equilibrium results for gapped graphene in the undoped limit. This comparison is facilitated by considering the case with impurities separated along the armchair direction for which the RKKY coupling is not complicated by the presence of short-range oscillations and long-range analytic results are available in the literature~\cite{GFR1,gapexch2,iop}. Without irradiation, graphene with a Dirac mass gap $\Delta_{\rm eq}$ can be characterized by the dimensionless parameter $\alpha_{{\rm eq}}=\Delta_{{\rm eq}} R/(2\hbar v_{F})$. For $\alpha_{{\rm eq}}\ll 1$, the leading-order RKKY coupling goes as  $1/R^{3}$~\cite{gapexch2} reducing to the case of gapless graphene~\cite{serami}. In contrast, under irradiation our results Eqs.~\eqref{exa3}-\eqref{exa4} show that even the leading term $\sim 1/R^{3}$ inherited from equilibrium gains a light-induced renormalization factor $F_0^2$. For $\alpha_{{\rm eq}}\gg 1$, the asymptotic large-$R$ behavior goes as  $e^{-2\alpha_{\rm eq}}/R^{3/2}$~\cite{GFR1,gapexch2}. To study the corresponding behavior in our case, we have extended our numerical calculation for the armchair case to larger values of $R$ for $\mathcal{A} = 0.06$, so that $\alpha_{\Delta} \gtrsim 1$.
  As shown in Fig.~\ref{fig8}, we find that our numerical results can be fitted to  $e^{-2\alpha_{\Delta}}/R^{3/2}$. The presence of the exponential decay factor as a function of $R$ is ubiquitous to indirect exchange interaction in insulators~\cite{Zhu_RKKY,BR_RKKY}.

\bibliography{refs_WK_v3}
\end{document}